\documentclass[twocolumn,secnumarabic,amssymb, nobibnotes, aps, prl, reprint,superscriptaddress]{revtex4-2}
\usepackage[english]{babel}
\usepackage{bm}
\usepackage{xcolor}
\usepackage{amsmath}
\usepackage{graphicx}
\usepackage{hyperref}
\usepackage{subcaption}
\usepackage{multirow}
\usepackage{tabularx}

\newcommand{\GW}{$G_0W_0$}
\newcommand{\gcn}{gC$_3$N$_4$}
\newcommand{\gcnt}{gC$_3$N$_4$-t}
\newcommand{\gcnh}{gC$_3$N$_4$-h}
\newcommand{\sqrtcell}{$\sqrt{3}\times\sqrt{3}$}
\newcommand{\atDFT}{@DFT}
\newcommand{\atGW}{@\GW}

\begin{document}
\title{First-Principles Calculations of Exciton Radiative Lifetimes in Monolayer Graphitic Carbon Nitride Nanosheets: Implications for Photocatalysis}

\author{Michele Re Fiorentin}%
\email{michele.refiorentin@iit.it}
\affiliation{Center for Sustainable Future Technologies, Istituto Italiano di Tecnologia, via Livorno 60, 10144, Torino, Italy}
\author{Francesca Risplendi}%
\affiliation{Dipartimento di Scienza Applicata e Tecnologia, Politecnico di Torino, corso Duca degli Abruzzi 24, 10129, Torino, Italy}
\author{Maurizia Palummo}
\affiliation{Dipartimento di Fisica and INFN, Universit\`{a} di Roma ``Tor Vergata'', via della Ricerca Scientifica 1, 00133, Roma, Italy}
\author{Giancarlo Cicero}%
\affiliation{Dipartimento di Scienza Applicata e Tecnologia, Politecnico di Torino, corso Duca degli Abruzzi 24, 10129, Torino, Italy}
\date{\today}%

\begin{abstract}
In this work, we report on the exciton radiative lifetimes of graphitic carbon nitride monolayers in the triazine- (\gcnt{}) and heptazine-based (\gcnh{}) forms, as obtained by means of ground- plus excited-state {\it ab~initio} calculations. By analysing the exciton fine structure, we highlight the presence of dark states and show that the photo-generated electron-hole pairs in \gcnh{} are remarkably long-lived, with an effective radiative lifetime of 260~ns. This fosters the employment of gC3N4-h in photocatalysis and makes it attractive for the emerging field of exciton devices. Although very long intrinsic radiative lifetimes are an important prerequisite for several applications, pristine carbon nitride nanosheets show very low quantum photo-conversion efficiency, mainly due to the lack of an efficient e-h separation mechanism. We then focus on a vertical heterostructure made of \gcnt{} and \gcnh{} layers which shows a type-II band alignment and looks promising for achieving net charge separation.
\end{abstract}

\maketitle

\noindent\textbf{Cite as:} \textit{ACS Appl. Nano Mater.} 2021, 4, 2, 1985–1993, \href{https://pubs.acs.org/doi/abs/10.1021/acsanm.0c03317}{doi/abs/10.1021/acsanm.0c03317}
\bigskip

After the seminal work of Wang et al. \cite{WANG2009} in 2009, proposing polymeric graphitic carbon nitride as a new, eco-friendly, low cost and thermally stable photocatalyst for 
hydrogen evolution, this layered material has become the subject of intense research efforts \cite{PC1,PC2,SiC1,C_photocatalysis,LIU841,C4GC01847H,C8SE00629F,Li:2020aa} for possible employment  
in several applications\cite{MONAI20181,C6GC02856J,Su:2010aa} as a ``green'' replacement of more expensive, polluting, metal-containing compounds \cite{Ramli:2018aa,C9QI00689C,Garino_2019,Risplendi_2020,juqin_Cu,fluffy_Cu}.  Among various allotropes of C$_3$N$_4$ with different density\cite{polymorphs,Fan:2016aa,SUN2019131}, 
the graphitic form has been proved to be the most stable at standard conditions \cite{graphitic_stability}. It is composed of two-dimensional (2d) layers of carbon and nitrogen atoms covalently bonded, stacked by means of van der Waals (vdW) interactions as in graphite. These weak dispersion forces allow to easily exfoliate graphitic carbon nitride \cite{exfoliation1,exfoliation2} into few-layers nanosheets or even single layer, graphene-like carbon nitride (\gcn). 
\gcn{} single layers can be built from either triazine or heptazine (tri-\textit{s}-triazine) molecules as basic units, forming, respectively, the so-called triazine- (\gcnt) or heptazine-based (\gcnh) graphene-like carbon nitride. 

Most of the experimental works on pristine graphitic carbon nitride investigate the optoelectronic properties and address the carrier dynamics with the goal of understanding the origin of the low photo-conversion efficiency \cite{C2EE03479D,Wang:2012aa,Melissen:2015aa,C7SC00307B}. Typical lifetimes of the photo-generated electron-hole pairs, measured in nanosheets with thickness ranging from few to several nanometers, are of the order of few nanoseconds, with values that change depending on the sample preparation and thickness\cite{niu2012,merschjann2013,dong2015,gan2016,giri2018,yuanjin2020}. 

On the theoretical side, a large number of {\it ab initio} studies, at different levels of approximation, have been carried out to clarify the structural, electronic and optical properties both of the bulk and the monolayer form \cite{gao,C7RA07134E,osorio-guillen,weiwei,steinmann,sun-yang}. Thanks to these studies, it is now clear that the most stable atomic structure is buckled\cite{gracia-kroll,azofra,deifallah-cora,buckling1x1} and that corrugation, by breaking the delocalised  $\pi$-bond, significantly alters the electronic and excitonic optical properties from those calculated for the flat, unstable structures. However, different corrugated geometries have been proposed and a complete description of the fundamental properties of both \gcnt{} and \gcnh{}, encompassing optical spectra, exciton fine structure and intrinsic radiative lifetimes is, to our knowledge, still lacking. 

Our aim is therefore to provide a thorough study of the electronic, optical and excitonic properties of both \gcnt{} and \gcnh{} in the buckled geometries without dynamical instabilities. This preliminary structural optimisation, performed using Density Functional Theory (DFT) simulations, enables us to reliably predict the quasi-particle (QP) electronic structures of both allotropes at the state-of-the-art $GW$ \cite{hedin,GW1,GW_BSE1} level. Our study shows that the effect of corrugation on the electronic properties is sizeable both at the DFT and GW level of approximation. While this behaviour has been already discussed in the literature, as we will point out in the following, our QP bandgaps present significant differences with previously published data.
The optical properties, obtained by solving the Bethe-Salpeter equation (BSE) \cite{bethe_salpeter,BSE1}, are characterised by pronounced excitonic effects associated to a rich structure of dark and bright strongly bound excitons, widely influenced by corrugation. 
Thanks to a recently derived approach \cite{Palummo:2015aa,PhysRevB.100.075135}, we compute, for both materials, the intrinsic exciton radiative lifetimes and their effective average, showing that \gcnh{} is characterised by a significantly long effective radiative lifetime. By suppressing electron-hole recombination, long exciton lifetimes can increase the chances of exciton dissociation and enhance the photocatalytic activity \cite{PhysRevLett.106.138302,catal3040942,C8TA04140G}. Moreover, excited states with long lifetimes can be particularly suitable for study and manipulation in external electromagnetic fields, as in the emerging domain of excitonic devices \cite{Miller:2017aa,Liueaba1830,Unuchek:2018aa,Uda:2016aa,cong2018excitons}.

To overcome the small availability of photo-generated free carriers, due to the large binding energies, exciton dissociation must be improved. Several mechanisms to achieve this goal have been proposed, ranging from doping with metal and non-metal atoms\cite{catal10101119,HASIJA2019494}, to order-disorder transitions \cite{Wang:2017ab}, to the creation of heterojunctions \cite{jingrun2018,wang2020,wang2020excitonic}. We show that a novel vdW heterostructure, in which a \gcnt{} and a \gcnh{} layer are vertically stacked, results in an interface with a type-II band alignment,  promising for obtaining net charge separation between the two layers.

In the quest for a deeper knowledge of this intriguing 2d material and for devising new applications in various fields, ranging from optoelectronics \cite{Wang:2017aa,electronics2,Grosso:10,andreakou2014}, to electrocatalysis \cite{azofra,Jin:2018aa,Younis:2019aa}, to photocatalysis\cite{TANG201930935,C8CY00970H,zhao-xu,KONG2020100488}, a detailed description and comparison of the electronic, optical and exciton properties is essential. Our results aim at providing such a resource, that can prove helpful in the understanding \gcn{} features and in the design of new technological applications.

The paper is organised as follows: in section \ref{sec:methods} we present the computational methods used to carry out our study, in section \ref{sec:results} we discuss the results and, finally, in section \ref{sec:conclusions} we draw the conclusions.
\begin{figure}[t]
\centering
\begin{subfigure}[b]{0.48\textwidth}
         \centering
         \includegraphics[width=\textwidth]{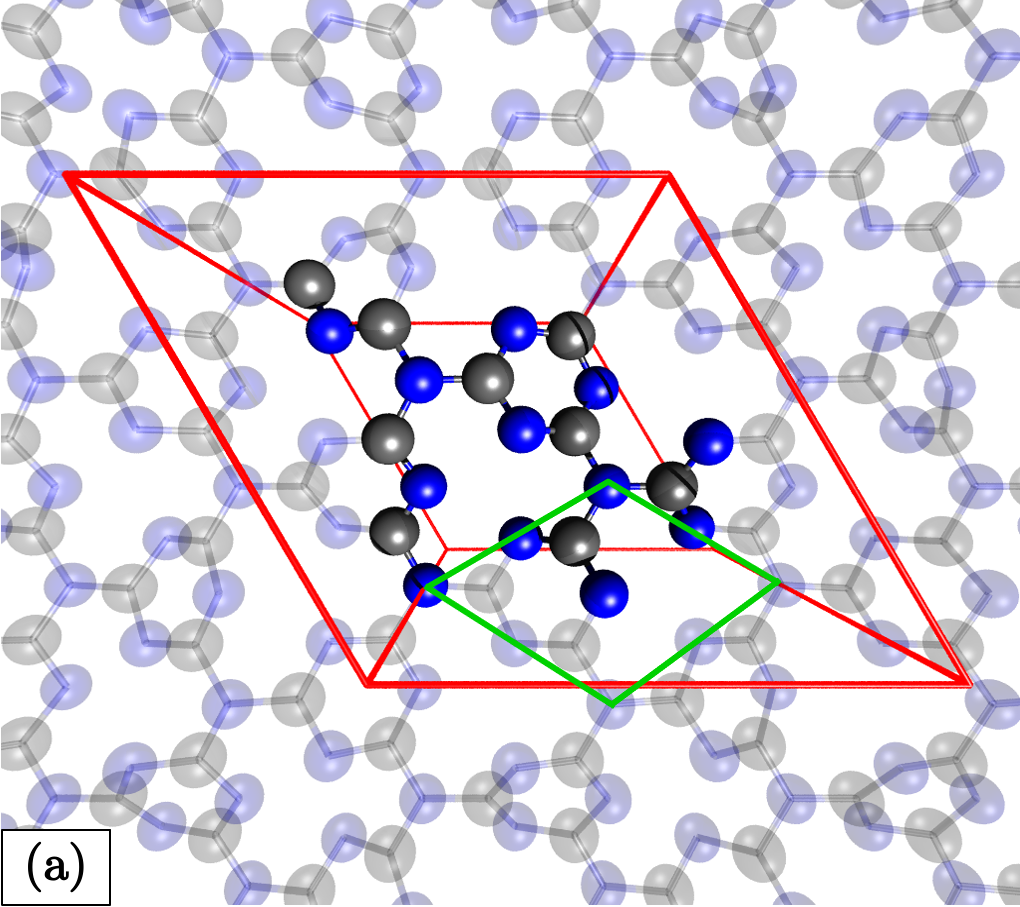}
\end{subfigure}
\hspace{.2cm}
\begin{subfigure}[b]{0.48\textwidth}
         \centering
         \includegraphics[width=\textwidth]{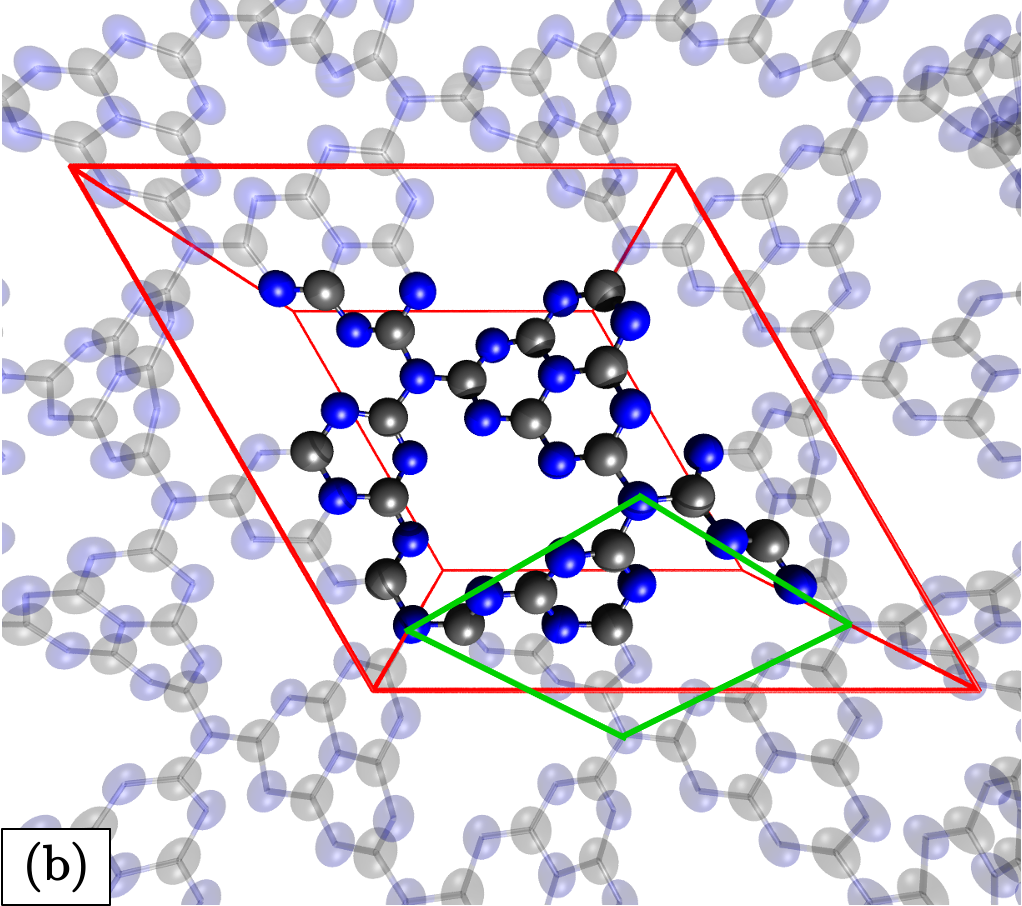}
\end{subfigure}
\caption{(a), (b) Perspective view of the optimised geometries of \gcnt{} and \gcnh{}, respectively. The \sqrtcell{} cells are marked with solid red lines, while the standard $1\times1$ cells with thinner, green lines. Carbon and nitrogen atoms are represented by grey and blue spheres, respectively.}
\label{fig:geometries}
\end{figure}
\section{Methods}
\label{sec:methods}
Structural optimisations are performed by means of DFT simulations carried out with the Quantum ESPRESSO package \cite{QE,QE2}. At this level of approximation, the gradient corrected Perdew-Burke-Ernzerhof (PBE) functional \cite{PBE} is employed to describe the exchange-correlation effects, and norm-conserving pseudopotentials \cite{ONCV} are used to model the electron-ion interactions. Wave functions are expanded in plane waves up to an energy cutoff of 80~Ry. To study various degrees of buckling, we employ different cells for both allotropes. We consider the standard unit cell, $1\times1$, of  \gcnt{} (\gcnh), comprising 7 (14) atoms (cf. Fig.~\ref{fig:geometries}), supercells obtained by its repetition along the lattice plane, such as $2\times2$ and $4\times4$ supercells, and rotated \sqrtcell{} cells, as shown in Fig.~\ref{fig:geometries}, containing 21 and 42 atoms for \gcnt{} and \gcnh{} respectively. In the case of the standard $1\times1$ cells, the Brillouin zone is sampled using a $4\times4\times1$ Monkhorst-Pack mesh \cite{MP} for \gcnt{} and a $2\times2\times1$ mesh for \gcnh{}. For the \sqrtcell{} cells, calculations are performed on a $2\times2\times1$ mesh for \gcnt{} and at the $\Gamma$-point for \gcnh. vdW interactions are treated within the semiempirical Grimme-D3 approach \cite{grimme-D3} and a vacuum region at least 15\AA{} thick along the direction orthogonal to the atomic layer is used to ensure the decoupling of the periodic replicas. 

Dynamical stability of the obtained structures is assessed through the calculation of their phonon Density of States (phDOS) \cite{dynamical_stability}. The presence of modes with imaginary frequencies is a sign that the considered structure, though an energy stationary point, tends to collapse to more stable geometries. phDOS are obtained with the PHONOPY code \cite{phonopy}, that employs a supercell, frozen-phonon approach. The dynamically stable, equilibrium geometries are then used for further calculations of their electronic and optical properties.

It is well known that DFT fails at properly describing the electronic structure and bandgap of semiconductors and insulators. To obtain precise results on the bandstructure and electronic gap, many-body perturbation theory techniques must be used. In this work, many-body corrections are calculated within the non self-consistent perturbative $GW$ (\GW) approximation, as implemented in the YAMBO code \cite{yambo1,yambo2}. The vacuum thickness is increased to 20~\AA{} and a box cutoff along the direction orthogonal to the lattice plane is employed. Bruneval-Gonze terminators \cite{bruneval-gonze} are used in the calculation of both the susceptibility and the correlation self-energy, and the plasmon-pole approximation is adopted. Convergence with respect to the free parameters of the theory is assumed when the value of the QP bandgap at the $\Gamma$-point varies less than 50~meV. Following this criterion, an $8\times8\times1$ ($6\times6\times1$) $k$-point mesh is used for \gcnt{} (\gcnh). 
We followed a ``ladder'' convergence method as laid out in Ref.~\cite{PhysRevMaterials.4.074009}, obtaining converged results with 400 (800) bands for \gcnt{} (\gcnh) and a dielectric matrix energy cutoff of 20~Ry for both allotropes.
More details on the convergence procedure are available in the Supporting Information.
\begin{figure}[t]
\centering
\begin{subfigure}[b]{0.44\textwidth}
         \centering
         \includegraphics[width=\textwidth]{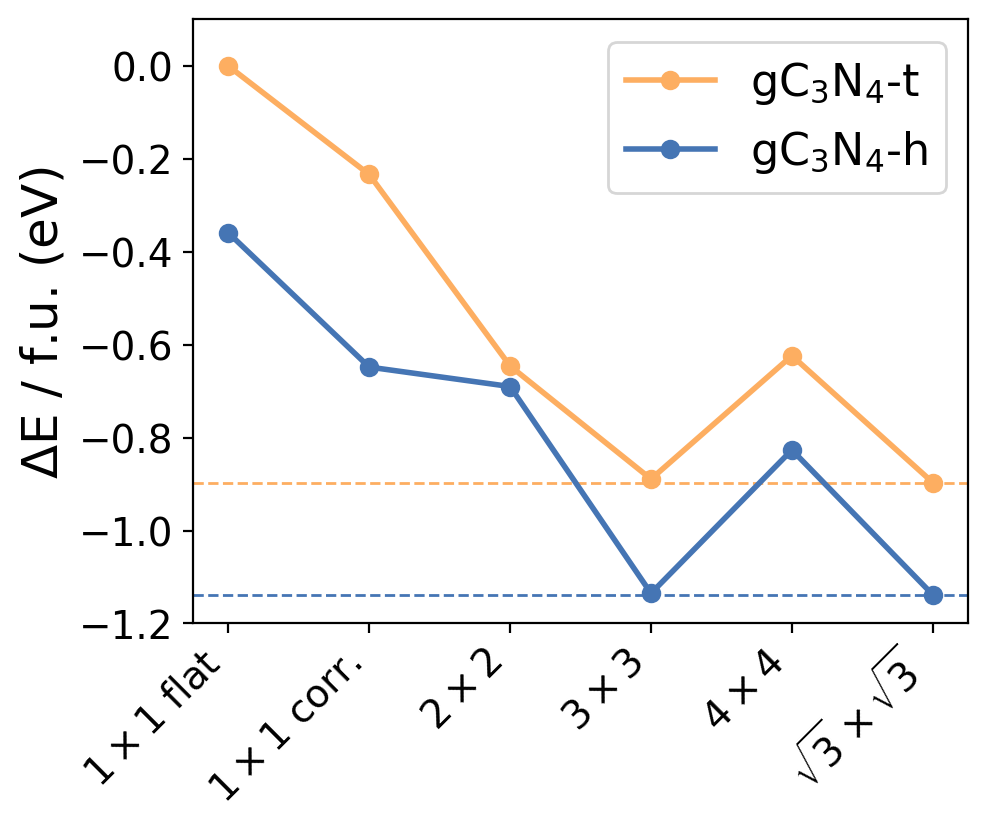}
         \caption{}
\end{subfigure}
\hspace{.4cm}
\begin{subfigure}[b]{0.45\textwidth}
         \centering
         \includegraphics[width=\textwidth]{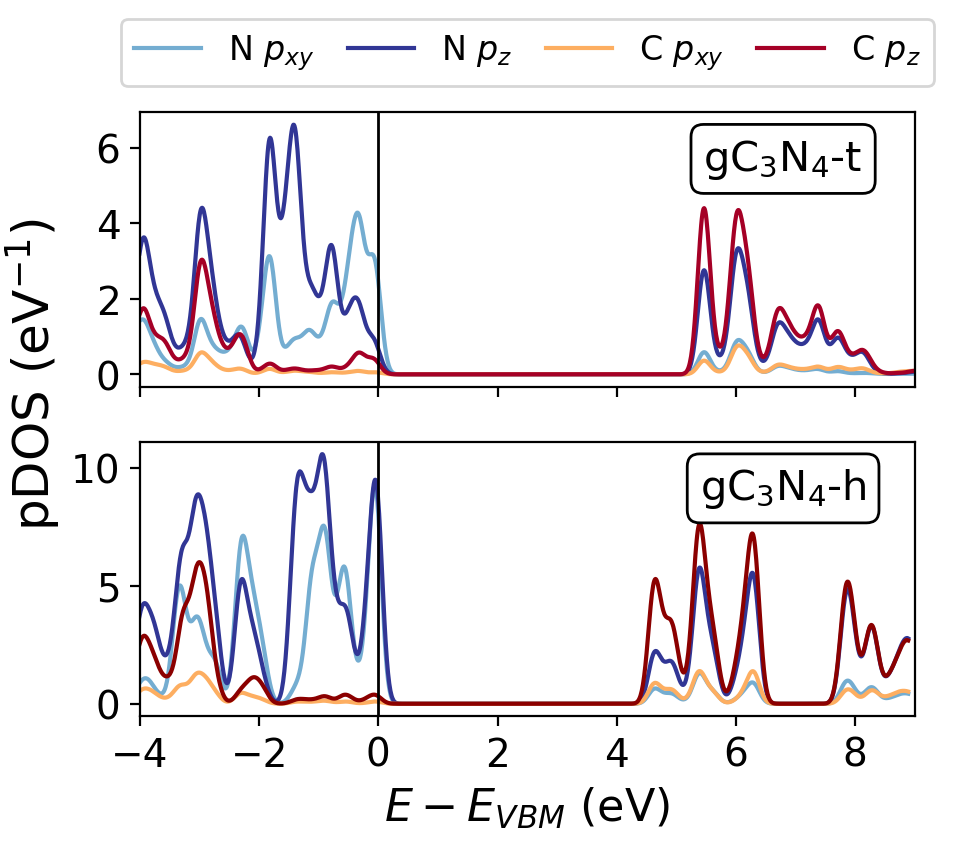}
         \caption{}
\end{subfigure}
\caption{(a) Energy per formula unit (f.u.) of \gcnt{} and \gcnh{} in different simulation cells, referenced to flat $1\times1$ triazine. (b) Projected Density of States of \gcnt{} (top panel) and \gcnh{} (bottom panel).}
\label{fig:en+pdos}
\end{figure}
Reliable optical properties can then be obtained by solving the BSE using the obtained QP corrections and adopting the Tamm-Dancoff approximation \cite{tamm,dancoff}. Converged spectra are computed within the static exchange approximation on denser $18\times18\times1$ and $9\times9\times1$ grids, using 150 and 300 bands for \gcnt{} and \gcnh, respectively, and a 2~Ry cutoff on the dielectric matrix for both materials. Once the BSE is solved, we compute the radiative lifetime, $\tau_S$, of each exciton state $S$ at $\mathbf{Q}=0$ using Fermi's golden rule \cite{PhysRevLett.95.247402,PhysRevB.100.075135,Palummo:2015aa}.
For Wannier excitons, the exciton dispersion in $\mathbf{Q}$ could be taken into account and the radiative lifetime renormalised by a thermal average on $\mathbf{Q}$, assuming a parabolic dispersion.
However, as shown in the following section, in our cases the bands involved in the relevant exciton states are almost flat in the neighbourhood of the $k$-points where the most significant transitions take place. This, in turn, causes a negligible exciton dispersion, which does not justify a thermal average over finite $\mathbf{Q}$ inside the light cone. Therefore, in what follows we shall employ the exciton radiative lifetimes calculated at $\mathbf{Q}=0$ and obtain the effective radiative lifetime, $\langle\tau\rangle_{\rm{eff}}$, at temperature $T$,  by assuming thermalisation and carrying out a Boltzmann average of the $\tau_S$ \cite{Perebeinos:2005aa}.

Finally, the unit cell of the vdW heterostructure made of a \gcnt{} and a \gcnh{} layer is obtained from the respective optimised \sqrtcell{} cells by means of the CellMatch code \cite{CellMatch}. 
\section{Results and discussion}
\label{sec:results}
We begin our study by comparing the stability of different \gcn{} structures built from the standard $1\times1$ cell and the \sqrtcell{} cell. The flat $1\times1$ cells are included in our study as references. When addressing buckling, the $1\times1$ cells show some level of corrugation, however, due to the periodic boundary conditions, they impose strong constraints on the atom positions. By considering larger supercells, such as $2\times2$, etc., artificial constraints on the geometry are gradually lifted, and the atomic layers acquire different degrees of corrugation. The buckling is due to the strong electrostatic repulsion between the lone pairs of the pyridinic-like nitrogen atoms that surround the void triangular regions in the atomic layer. 
Lifting geometrical constraints allows for a certain degree of relaxation so that the atoms can rearrange in such a way that opposite lone pairs do not directly face each other. Nevertheless, due to the periodic boundary conditions imposed by the computational scheme, geometrical artefacts can be fully avoided either in the limit of very large supercells or by identifying a reasonably small supercell that proves to be an energetic minimum. 
Pursuing the second strategy, in Fig.~\ref{fig:en+pdos} (a) we report the relative energy per formula unit (f.u.)\ of both \gcnt{} and \gcnh{}, studied in different simulation cells, with respect to an f.u.\ of flat \gcnt{}. The cell parameters and the relative energies per f.u.\ are reported in the Supporting information. It is evident that by enlarging the standard unit cell an energy minimum is reached with the $3\times3$ supercell. Further expanding the supercell to $4\times4$ introduces other artificial constraints on the buckling periodicity that result in a higher energy. The $3\times3$ supercells hence appear to provide a fully relaxed geometry, however, it is possible to notice that the smaller, rotated, \sqrtcell{} cells reported in Fig.~\ref{fig:geometries} show the same energy per f.u.\ and buckling pattern. The phDOS computed for \gcnt{} and \gcnh{} in the flat $1\times1$ and \sqrtcell{} cells, (c.f. Supporting information), demonstrate that, while the flat geometry is dynamically unstable, the \sqrtcell{} structures do not show soft phonon modes. 
We also note that for all considered simulation cells, the heptazine-based allotrope presents a lower energy per formula unit than the corresponding \gcnt{} structure, though the stabilisation of \gcnh{} with respect to \gcnt{}, both in \sqrtcell{} geometry, is less than 40~meV/atom. We shall then adopt as equilibrium geometries the \sqrtcell{} structures and compute the electronic and optical properties for both allotropes. 

\begin{figure}
\centering
\includegraphics[width=\linewidth]{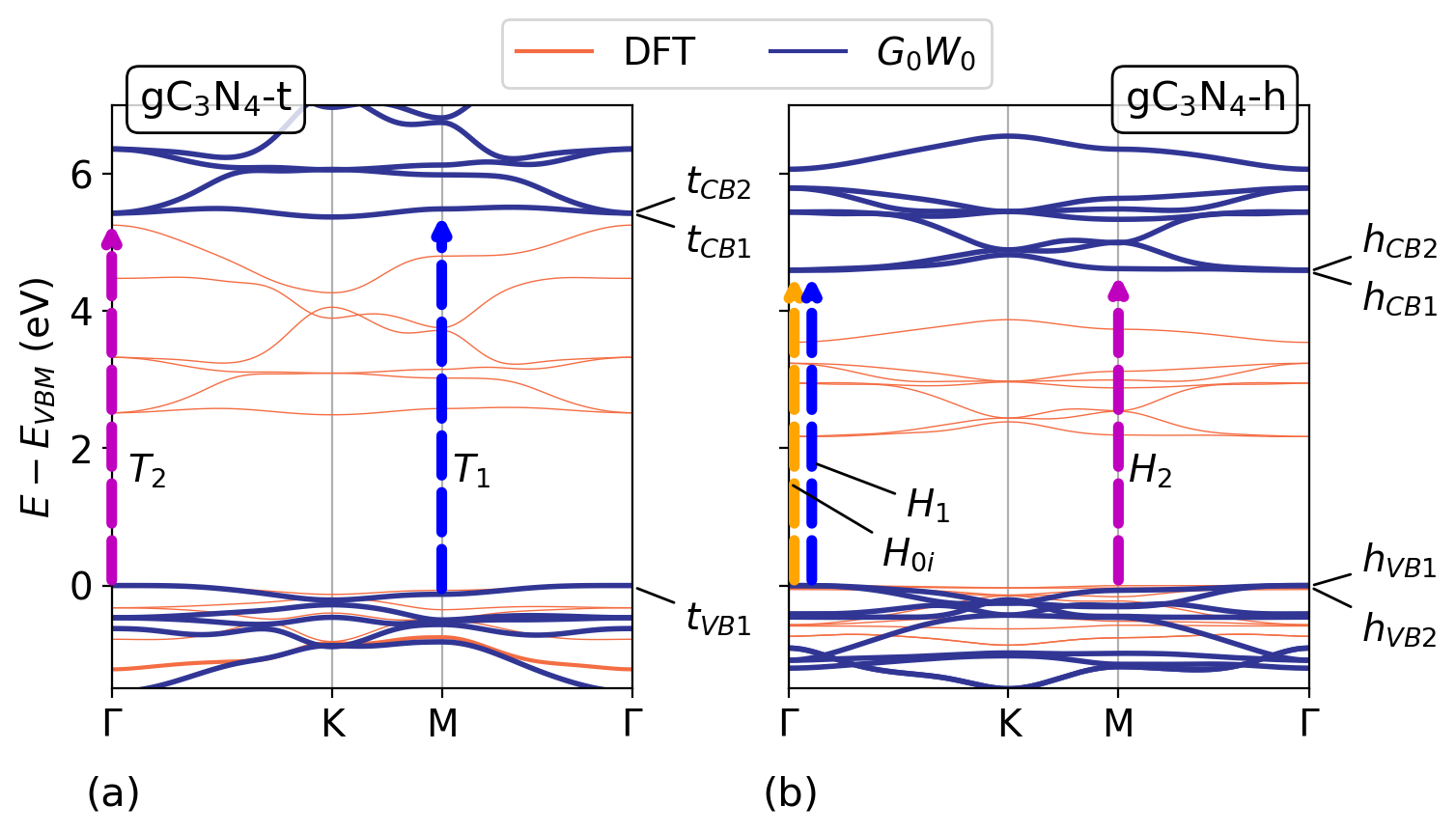}
\caption{Band structures of \sqrtcell{} \gcnt{} (a) and \gcnh{} (b), computed at the DFT (thin red lines) and at the \GW{} (blue lines) level of approximation. The dashed vertical arrows mark the electronic transitions contributing to some exciton states obtained by solving the BSE. The arrows in panel (a) and (b) correspond to the crosses in the upper panels of Fig.~\ref{fig:bse} (a) and (b), respectively.}
\label{fig:QP}
\end{figure}

In Fig.~\ref{fig:QP} we report the electronic bandstructure of \gcnt{} (a) and \gcnh{} (b) computed in their \sqrtcell{} cell, \atDFT and \atGW.

\gcnt{} presents an indirect electronic gap between the valence band maximum, located at the $\Gamma$ point, and the conduction band minimum at the $K$ point. The indirect gap is computed to be 2.48~eV \atDFT{} and is then widened to 5.37~eV \atGW. The direct gap at the $\Gamma$ point is only slightly larger, measuring 2.52~eV \atDFT{} and 5.42~eV \atGW. It is possible to notice that the effect of corrugation and stabilisation of the structure leads to important changes to the electronic properties. In particular, the stress minimisation of the \sqrtcell{} cell results in a slightly indirect gap, as opposed to previous results in the literature \cite{weiwei,steinmann}. It is also worth pointing out that the direct gap obtained from the optimised, buckled structure is more than 1~eV larger than the result for flat geometry \cite{weiwei}, both \atDFT{} and \atGW.

\gcnh{} displays a direct electronic gap at the $\Gamma$ point, 2.17~eV wide \atDFT{} and 4.60~eV \atGW. We note that the QP bandgap of the flat structure was, instead, computed to be indirect, 4.15~eV wide, while the direct bandgap was 2.09~eV \atDFT{} and 5.22~eV \atGW \cite{weiwei}.

From Fig.~\ref{fig:QP} (a) and (b) it is evident that the electronic structures of both \gcnt{} and \gcnh{}, in the \sqrtcell{} buckled geometry, feature bands that show little dispersion along the Brillouin zone path. This reflects the higher spatial localisation of the electronic states imposed by corrugation and, in particular, the breaking of the delocalised $\pi$ bond that characterises the lower conduction band in the flat geometry. When corrugation is considered, the $p_z$ orbitals of C and N, that contribute to the lower conduction band as seen in Fig.~\ref{fig:en+pdos}(b), cannot create a complete, delocalised $\pi$-system across the structure, but are forced to form much more localised and less stabilised states. The impact of buckling on the electronic band structure can then be regarded as twofold: on the one hand, the destabilisation of the states contributed by carbon and nitrogen $p_z$ orbitals, i.e. the lower conduction bands of both \gcnt{} and \gcnh{}, causes the opening of the bandgap already at the DFT level. On the other hand, a higher electronic localisation increases the screening and results in a gap opening due to the \GW{} corrections which tends to be smaller than what observed in the flat case.

The QP energies computed \atGW{} are then employed in the calculation of the optical absorption of both allotropes. 
\begin{figure}
\centering
\includegraphics[width=\linewidth]{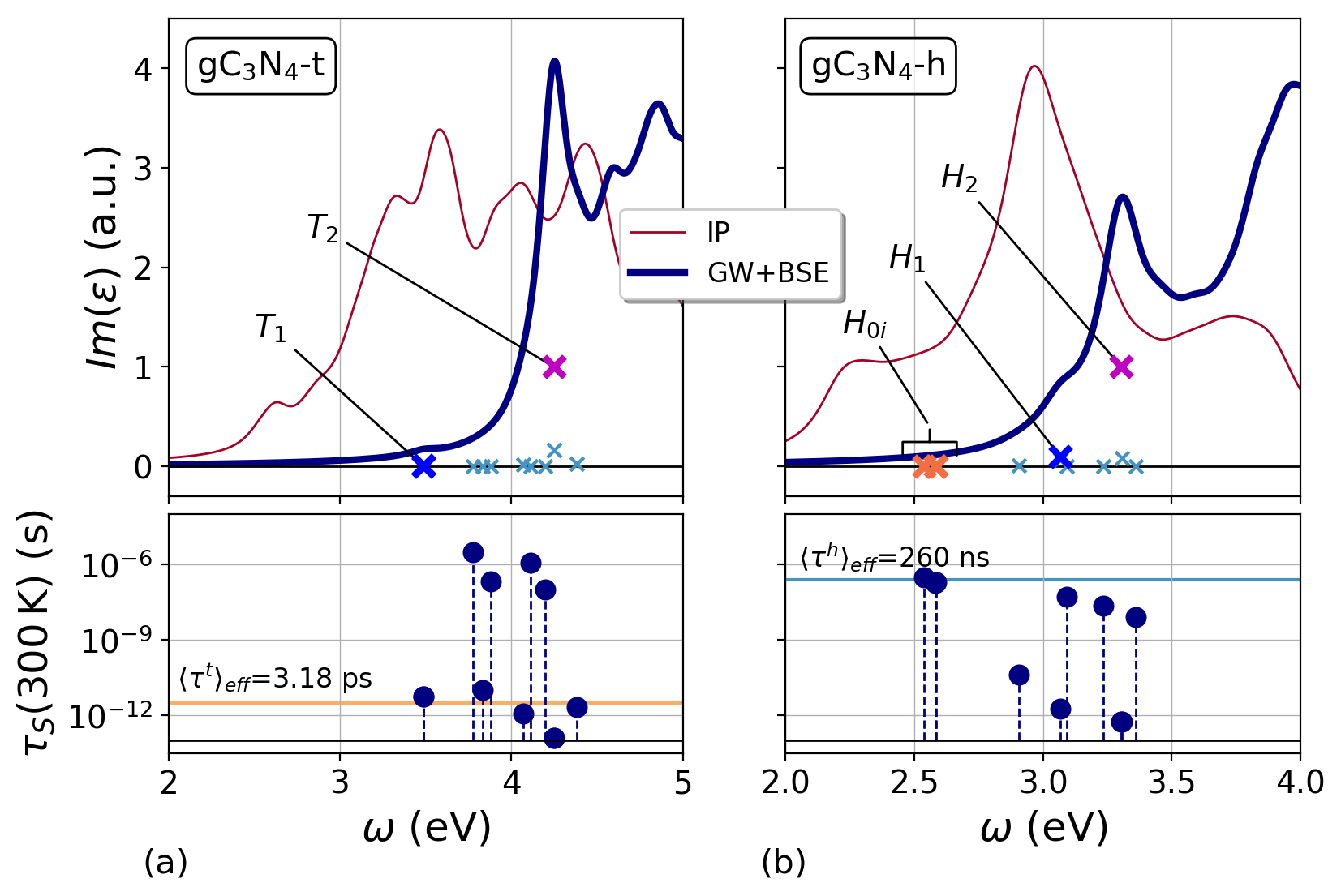}
\caption{(a), (b) upper panels: imaginary part of the macroscopic dielectric function computed for \gcnt{} and \gcnh{}, respectively, at the independent particle and GW+BSE levels. The crosses mark the excitonic states obtained from the BSE. (a), (b) lower panels: exciton radiative lifetimes of \gcnt{} and \gcnh{}, respectively, at $T=300$~K. The horizontal lines mark the effective radiative lifetimes $\langle\tau\rangle_{\rm{eff}}$.}
\label{fig:bse}
\end{figure}
In the upper panels of Fig.~\ref{fig:bse}(a) and (b) we report the imaginary part of the macroscopic dielectric function of \gcnt{} and \gcnh{}, respectively, computed with different approaches. The thin red lines show the spectra obtained with the independent-particle (IP) random phase approximation based on the PBE energy levels. The thick blue lines represent the spectra obtained by solving the BSE built by using the QP energies. It is evident that the IP approach, using PBE energetics, largely redshifts the spectra. On the contrary, the solution of the BSE allows to obtain reliable absorption spectra by taking into account excitonic effects.

In the upper panel of Fig.~\ref{fig:bse} (a) it is possible to notice that the optical absorption of \gcnt{} shows a pronounced peak at 4.25~eV, related to the $T_2$  bright exciton. The optical gap, however, is given by the bright $T_1$ exciton, which also corresponds to the lowest-energy solution of the BSE, and is located at 3.49~eV. By analysing the exciton wavefunctions, it is possible to identify which electronic transitions give the largest contributions. In particular, as marked by the arrows in Fig.~\ref{fig:QP} (a), $T_2$ receives significant contributions from transitions around the $\Gamma$ point between the top valence band (tVB) and the degenerate bottom conduction band (bCB), while $T_1$ is mostly given by transitions around the $M$ point between the tVB and the non-degenerate bCB. Although the electronic gaps at $\Gamma$ and at $M$ differ by less than 0.2~eV, the binding energies of $T_1$ and $T_2$ are $E_{T_1}^b=1.94$~eV and $E_{T_2}^b=1.18$~eV. Indeed, as shown in Fig.~\ref{fig:QP} (a), at the $\Gamma$ point, the bCB is made of two degenerate bands, $t_{\rm{CB1}}$ and $t_{\rm{CB2}}$, one of which, $t_{\rm{CB2}}$, shows a larger dispersion along the $\overline{\Gamma K}$ direction, while, at $M$, the bCB is only composed of the almost flat $t_{\rm{CB1}}$ band. Therefore, around $\Gamma$ the electronic transitions involve less localised states which, in turn, yield a smaller screening than more localised states as $t_{\rm{CB1}}$ around $M$. A smaller screening implies a reduced direct electron-hole interaction and thus a smaller exciton binding energy, so that $E_{T_1}^b>E_{T_2}^b$. In the lower panel of Fig.~\ref{fig:bse} (a) we report the exciton radiative lifetimes of the exciton states between $T_1$ and $T_2$. Here we can notice the presence of several dark excitons, with very small dipole strength, with $\tau_S\sim 1\,\mu\rm{s}$, while $\tau_{T_2}=0.1$~ps and $\tau_{T_1}=3.1$~ps. The thermal average favours the lowest energy states, so that the effective radiative lifetime of \gcnt{} is dictated by the lifetime of $T_1$ and, at $T=300$~K, results in $\langle\tau^t\rangle_{\rm{eff}}=3.18$~ps.

The absorption spectrum and excitonic properties of \gcnh{}, upper panel of Fig.~\ref{fig:bse} (b), look richer. In the lowest part of the spectrum we notice the presence of a closely spaced triplet of excitons $H_{0i}$, with $i=a,b,c$. These are all dark and the lowest-energy exciton, $H_{0a}$, is located at 2.57~eV. As shown in Fig.~\ref{fig:QP} (b), the tVB of \gcnh{} around the $\Gamma$ point is composed of two degenerate bands, $h_{\rm{VB1}}$ and $h_{\rm{VB2}}$ and, likewise, the bCB is given by the degenerate $h_{\rm{CB1}}$ and $h_{\rm{CB2}}$ bands. The dark excitons $H_{0i}$ are mainly composed of transitions around the $\Gamma$ point involving $h_{\rm{VB1}}$ and $h_{\rm{CB1}}$, which are dipole-forbidden. 
The first bright exciton, responsible for a noticeable bump in the spectrum, is $H_1$ at 3.08~eV, mainly due to transitions between $h_{\rm{VB1}}$ and $h_{\rm{CB2}}$ around $\Gamma$, while at 3.30~eV we find $H_2$, which forms the first strong absorption peak and receives the largest contributions from transitions between $h_{\rm{VB1}}$ and $h_{\rm{CB1}}$ around the $M$ point.\\
Some comments on this complex exciton pattern are in order. Excitons $H_{0i}$ and $H_1$ receive contributions from transitions between bands that are degenerate in energy, nonetheless their binding energies are largely different, e.g. ${E^b_{H_{0a}}=2.03}$~eV and $E^b_{H_1}=1.52$~eV. This discrepancy can again be traced back to differences in the direct electron-hole interaction in the two cases. 
As shown Fig.~\ref{fig:en+pdos}(b), the tVB of \gcnh{} receives significant contributions from N $p_z$ orbitals, and so does the $h_{\rm{VB1}}$ state at the $\Gamma$ point ($h_{\rm{VB1}}@\Gamma$). bCB states $h_{\rm{CB1}}@\Gamma$ and $h_{\rm{CB2}}@\Gamma$ both result from C and N $p_z$ orbitals, but their spatial localisations differ from each other (cf. Supporting Information). For this reason, the overlap between $h_{\rm{VB1}}@\Gamma$ and $h_{\rm{CB1}}@\Gamma$ is large and screening plays a significant role, giving the particularly large exciton binding energy of $H_{0i}$, while the overlap between $h_{\rm{VB1}}@\Gamma$ and $h_{\rm{CB2}}@\Gamma$ is small, so that  $E^b_{H_{0i}}>E^b_{H_1}$. Similar considerations can be drawn for $H_2$ (cf. Supporting Information), given that the electronic bandgap at $M$ differs from the gap at $\Gamma$ only by some meV, while $E^b_{H_{0i}}> E^b_{H_3}=1.51$~eV. 

It is important to notice that, as evident from Fig.~\ref{fig:bse} (b), the optical absorption edge, around $H_1$, falls in the blue-violet part of the visible spectrum, as opposed to the \gcnt{} case, cf. Fig.~\ref{fig:bse} (a), in which the absorption onset, marked by $T_1$, is already in the near ultraviolet. The absorption edge value of \gcnh{} obtained in this work, $\sim 3$~eV, is in line with the experimental optical gap measurement of $\sim2.7$~eV \cite{exp_opt_gap,exp_opt_gap1}, provided one considers that the absorption measurements are carried out on layered systems. Indeed, a larger optical gap is expected in single-layer 2d samples, as in our study, in comparison to bulk or few-layer materials such those experimentally explored \cite{C7RA07134E}. 

In the lower panel of Fig.~\ref{fig:bse} (b) we report the radiative lifetimes of the exciton states up to around $H_2$. As in the case of \gcnt{}, we can notice several dark states, however the crucial property of \gcnh{} is the presence of the three dark excitons $H_{0i}$ as lowest-energy states. Their lifetimes are $\tau_{H_{0i}}\simeq 200$~ns, while ${\tau_{H_2}=0.5}$~ps. The thermal average is dominated by the lifetimes of the lowest-energy states, therefore, at $T=300$~K, we obtain $\langle\tau^h\rangle=260$~ns. Since the bottom of the \gcnh{} exciton spectrum is composed of dark states, when thermalisation is assumed, these long-lived levels become highly populated and the system remains in an excited state for a rather long period of time. Heptazine-based graphitic carbon nitride has indeed shown remarkably long exciton lifetimes in various experiments, mainly involving the nanosheet form. In Ref. \cite{dong2015}, by means of time resolved measurements of decay spectra, Dong et al.\ obtained an average lifetime of the photo-generated electrons and holes of around 37~ns in nanosheets 16 nm thick. Similar results were collected by Choudhury et al., which, in Ref. \cite{giri2018}, measured an average lifetime of about 20~ns in graphitic C$_3$N$_4$ nanosheets composed of more than 10 atomic layers. Niu et al.\ pointed out in Ref.~\cite{niu2012} that the average exciton lifetime in graphitic C$_3$N$_4$ decreases when increasing the thickness of the sample due to quantum confinement effects, which become stronger when dimensionality is reduced.
Therefore, by considering the differences in thickness and defectivity between our graphene-like monatomic layer and the experimentally studied few-layer nanosheets, we argue that our results can provide a reliable theoretical prediction on the order of magnitude of the effective exciton radiative lifetime in \gcnh{}.

\begin{figure}[t]
\centering
\begin{subfigure}[b]{0.5\textwidth}
         \centering
         \includegraphics[width=\textwidth]{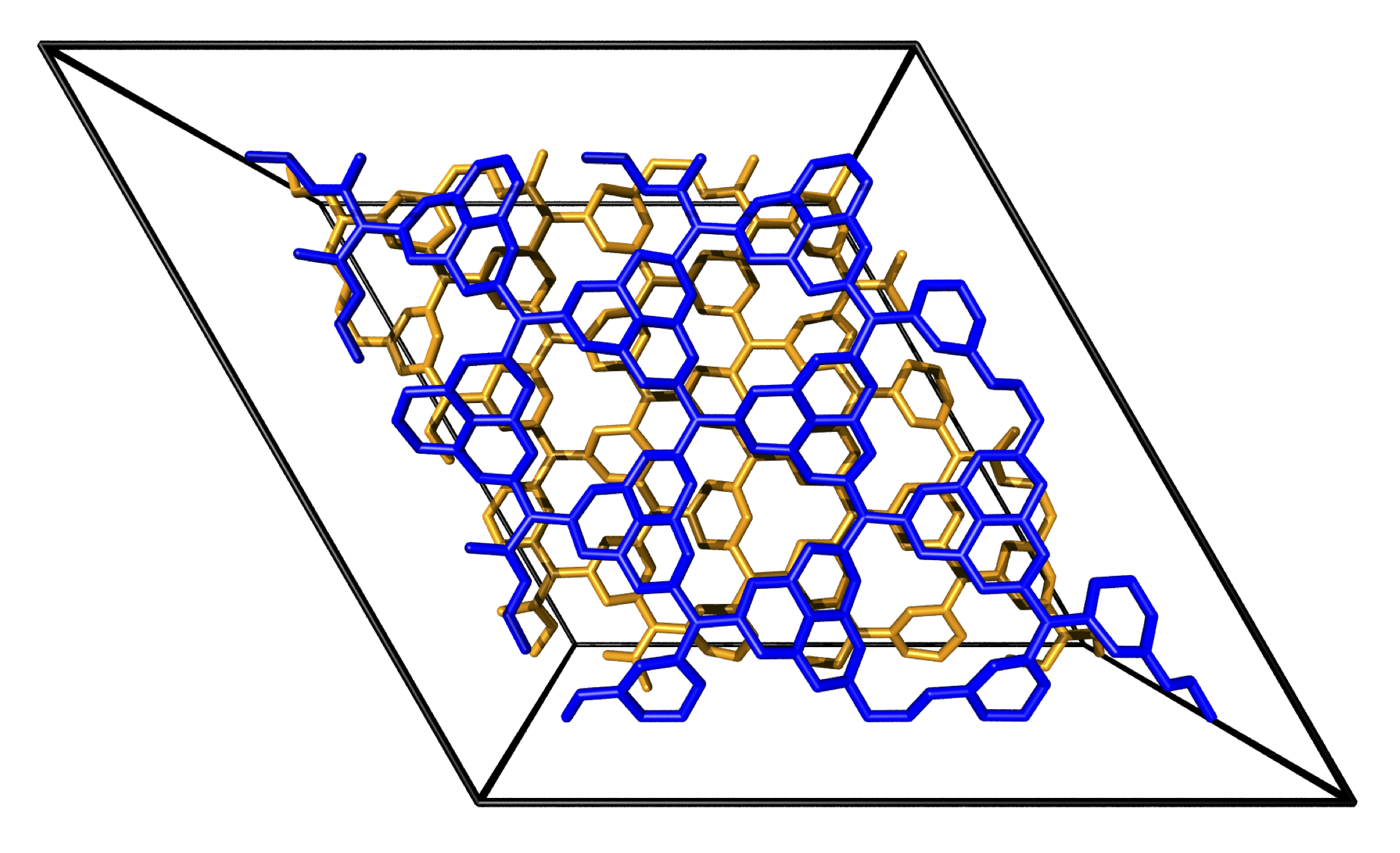}
         \caption{}
\end{subfigure}
\begin{subfigure}[b]{0.4\textwidth}
         \centering
         \includegraphics[width=\textwidth]{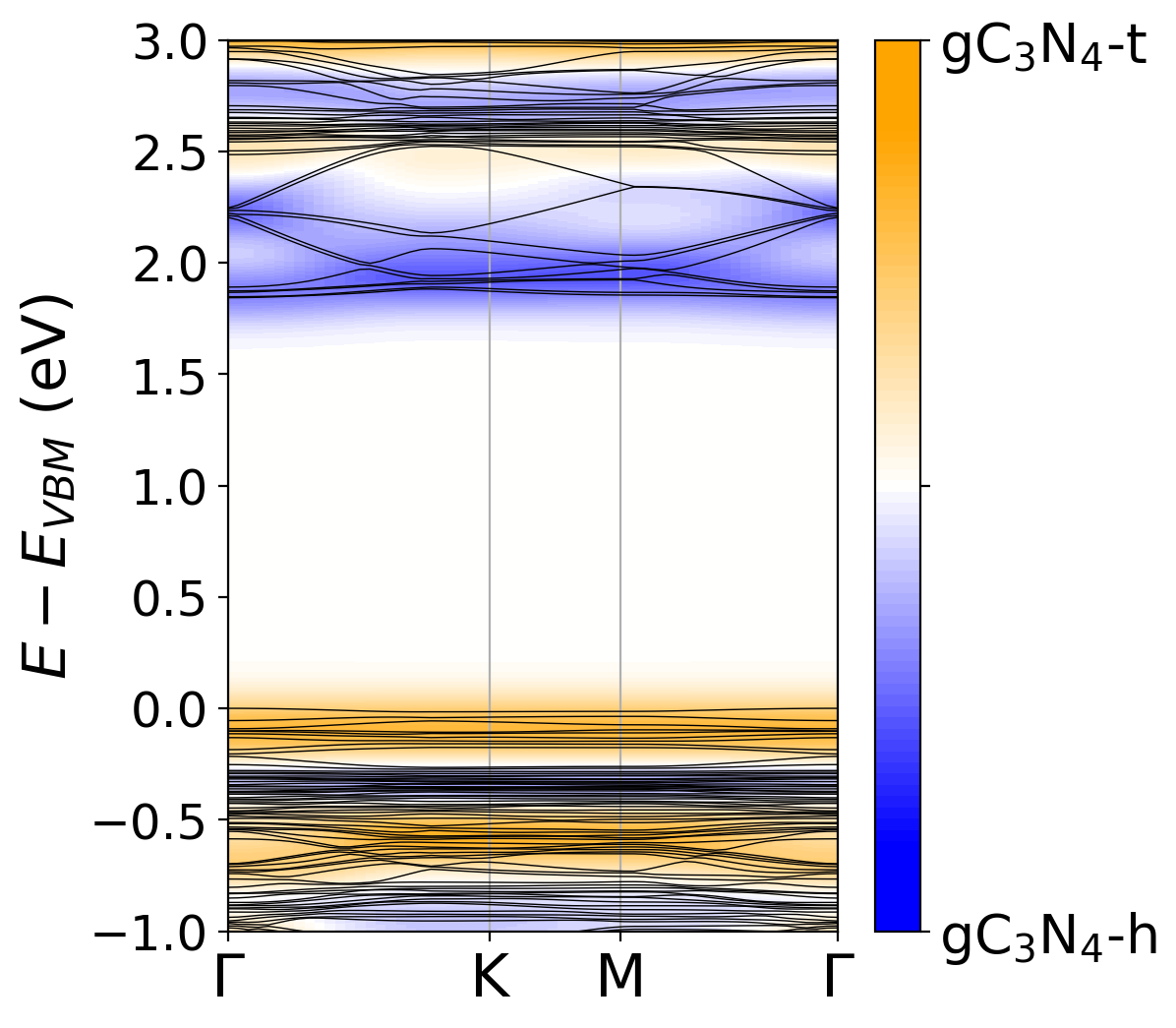}
         \caption{}
\end{subfigure}
\caption{(a) Perspective view of the \gcnt{}/\gcnh{} vdW heterostructure. The \gcnt{} and \gcnh{} layers are represented by orange and blue segments, respectively. The periodic cell is marked by the black lines. (b) Heterostructure electronic bands and $k$-resolved DOS. Bands are colored according to the layer whose atomic orbitals contribute the most. The blue color marks contributions from \gcnh{} atoms only, white equal contributions from both layers, while orange signals contributions from \gcnt{} atoms only. 
}
\label{fig:interface}
\end{figure}

Finally, aiming at suggesting possible strategies to favour exciton dissociation, we focus on a vertical vdW heterostructure made of \gcnt{} and \gcnh{} layers. The heterostructure relaxed geometry is reported in Fig.~\ref{fig:interface} (a), where the \gcnt{} and \gcnh{} layers are represented by the orange and blue nets respectively. 
The strain along the lattice vectors is around 1.5\% with respect to the equilibrium lattice parameters of the single layers and the cohesion energy is -24~meV/\AA$^2$, with respect to the separate layers. The intriguing features of the heterostructure come to light with the analysis of its electronic bandstructure. In Fig.~\ref{fig:interface} (b) we show the bandstructure computed \atDFT. Though affected by the notorious limitations, this approach is more than sufficient for the purpose of gaining a qualitative insight on the heterostructure properties. To the band plot, we superimpose the $k$-resolved DOS projected on the orbitals of the atoms composing the two layers. This way, the bands along the $k$ path are coloured according to the difference between the DOS contributions from the two layers, allowing us to identify whether the electronic states at various $k$ points are localised on the \gcnt{} layer, or on the \gcnh{} layer, or on both. The heterostructure bandgap is 1.84~eV \atDFT{}, smaller than both \gcnt{} and \gcnh{} single layers, showing the effect of the interaction between the two layers. Moreover, it is evident from Fig.~\ref{fig:interface} (b) that the upper valence band is given by states localised on the \gcnt{} layer, while the lower conduction band appears localised on \gcnh{}, realising a type-II band alignment. Although a detailed study of different stacking patterns is needed to pinpoint the most stable heterostructure geometry, we expect that the main feature of a type-II band alignment will not be affected, due to the weak interaction between the two layers. It is then conceivable that either a direct transition between tVB and bCB occurs, leaving the hole on the \gcnt{} layer and promoting the electron on \gcnh{}, or, should the dipole strength of the direct transition vanish, an indirect, phonon-mediated, mechanism takes place, leading to the same final result  \cite{malic2,tomanek}. Indeed, when considering vdW heterostructures, transitions between states belonging to different layers are very often suppressed due to the scarce overlap of the relative electronic states. However, even in the case of optical transitions taking place between two \gcnh{} states, the type-II band alignment is such that the system is at its energetic minimum only when the hole occupies the valence band maximum, located on \gcnt{}. It has been shown \cite{kira,malic1} that this final state can be efficiently reached through tunnelling and the emission of phonons with suitable energy and momentum, which make the system relax to its energy minimum, thus enhancing the exciton lifetime and charge separation~\cite{malic3}.
\section{Conclusions}
\label{sec:conclusions}
In this work we have presented a thorough analysis of various properties of the increasingly widespread two-dimensional graphene-like carbon nitride. The features that make this material attractive for new technological applications are mainly related to its optical absorption and exciton fine structure. These, in turn, are extremely sensitive to the geometry, dimensionality and strain of the material. For these reasons, a stability analysis of different corrugation geometries has been carried out, highlighting the fundamental role of buckling and pointing at the \sqrtcell{} cell as the most stable geometry for both allotropes. 

Only once these stable structures have been identified it is possible to reliably discuss the electronic properties at the \GW{} level, identifying rather large bandgaps of the order of 5~eV. In particular, the bandgap of \gcnt{} in our buckled \sqrtcell{} cell turns out slightly indirect and larger than the direct bandgap of \gcnh{}. The QP levels have allowed us to compute the optical absorption spectra of both allotropes by solving the BSE and revealing their exciton patterns. In both cases large exciton binding energies have been computed, in line with the existing literature. While \gcnt{} shows an absorption edge in the near UV region and no low-energy dark excitons, the absorption edge of \gcnh{} falls in the blue-violet part of the visible spectrum and dark excitons are present at a lower energy. These properties directly impact on the viability of \gcnh{} in optoelectronics, photovoltaics and photocatalysis. Although absorption occurs in an energy window in which the intensity of the solar spectrum is modest, the presence of dark excitons as lowest-energy states largely affects the average radiative lifetime, lifting it to 260~ns. This long excitation lifetime dramatically increases the probability that the photo-generated electrons and holes contribute to photocatalytic reactions before recombination. In addition, single layer \gcnh{} may represent an interesting material in optoelectronics, appearing particularly promising from the perspective of exciton devices, in which long-lived excitons are required. 

Finally, we have proposed a novel vdW heterostructure containing one layer of each allotrope. We have predicted a type-II band alignment between the two materials, with the heterostructure tVB consisting of electronic states localised on \gcnt{} and the bCB localised on \gcnh{}. This appears particularly promising for enhanced charge separation between the two layers, achieved either by direct photo-excitation or through the polarisation-to-population mechanism \cite{malic1}. 

In conclusion, these results shed light on this promising bidimensional material, and can serve as a guide to understanding \gcn{} peculiar properties and designing novel applications that best exploit its predicted features.
\subsection{acknowledgement}
The authors acknowledge CINECA for the availability of high performance computing resources under the Iscra-B initiative, as well as the computational support provided by HPC@POLITO (http://www.hpc.polito.it).

\bibliographystyle{science}
\bibliography{gC3N4}

\clearpage
\onecolumngrid
\section{Supporting Information}

\subsubsection*{Convergence details of \GW{} and BSE calculations}
\begin{figure}[h]
\centering
\begin{subfigure}[b]{0.3\textwidth}
         \centering
         \includegraphics[width=\textwidth]{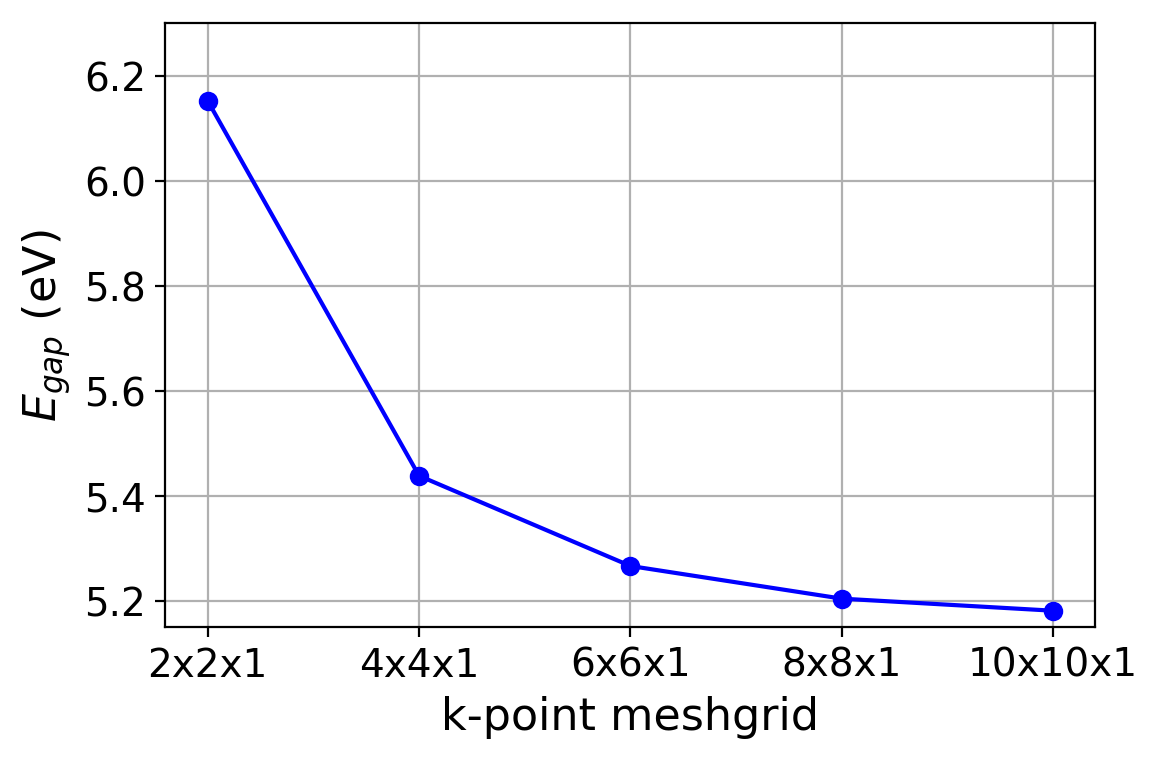}
         \caption{}
         \label{fig:gcnt_kpts}
\end{subfigure}
\hspace{0.2cm}
\begin{subfigure}[b]{0.3\textwidth}
         \centering
         \includegraphics[width=\textwidth]{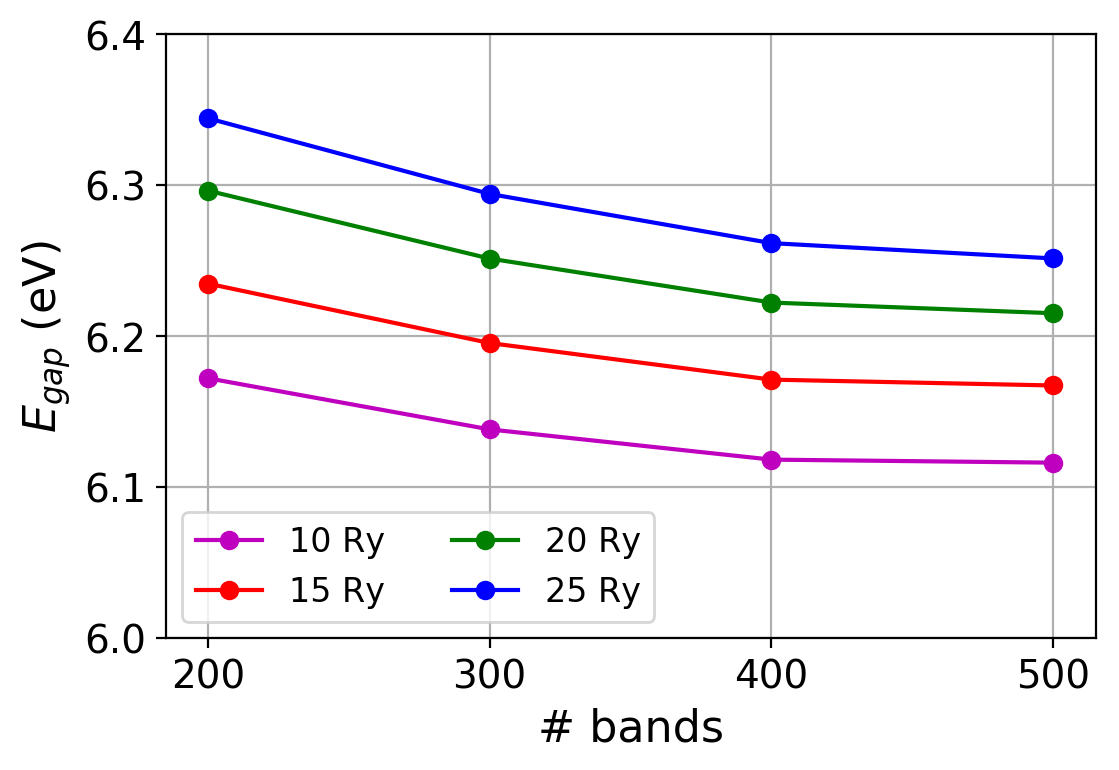}
         \caption{}
         \label{fig:gcnt_bnds_blks}
\end{subfigure}
\caption{Convergence tests on \GW{} calculations. (a) Variation of the QP bandgap at the $\Gamma$ point with the $k$-point grid. (b) QP bandgap at the $\Gamma$ point obtained on the coarse $2\times2\times1$ grid by varying the number of bands and the energy cutoff of the dielectric screening.}
\end{figure}

To determine the convergence parameters of our \GW{} calculations, we followed a ``ladder'' approach. Adopting the quasiparticle (QP) bandgap at the $\Gamma$ point as convergence parameter and a 50~meV threshold, we first performed a test on the $k$-point grid. In the case of \gcnt{}, a $8\times 8\times 1$ grid was chosen, Fig.~\ref{fig:gcnt_kpts}.
Convergence tests on the dielectric screening $\varepsilon_\mathbf{GG'}$ and on the correlation self energy $\Sigma_c$ were carried out on the coarse $2\times2\times1$ $k$-grid, given their very weak dependence on the Brillouin zone sampling. In Fig.~\ref{fig:gcnt_bnds_blks} we report the results of calculations on \gcnt{} obtained by varying both the number of bands $N_b$ included in the calculation of the screening and the energy cutoff $E_\mathbf{G}$ on $\varepsilon_\mathbf{GG'}$. $N_b=400$ and $E_\mathbf{G}=20$~Ry are sufficient for reaching convergence within the adopted threshold. Lower parameters $N'_b=400$ and $E'_\mathbf{G}=20$~Ry were instead used to perform a full calculation on the denser $8\times8\times1$ grid. The obtained QP levels were then corrected with a rigid scissor operator computed on the coarse $2\times2\times1$ grid. This was obtained by computing the difference between the QP bandgaps calculated at the convergence parameters $N_b$, $E_\mathbf{G}$ and at $N'_b$, $E'_\mathbf{G}$. This method helps reducing the computational effort on systems with many atoms, but clearly introduces a systematic error by correcting the QP energies at different $k$-points with the same rigid shift obtained at $\Gamma$. However, by means of a fully-fledged benchmark calculation, we checked that the error due to this ``ladder'' convergence method lies always within the fixed threshold of 50~meV. Finally, the usage of terminators in the calculation of $\Sigma_c$ allowed for the employment of the same number of bands $N_b$ used for the dielectric screening. 

In the case of \gcnh{}, converged results were obtained with the same method, using a dense $6\times6\times1$ and a coarse $2\times2\times1$ $k$-grids, $N_b=800$, $E_\mathbf{G}=20$~Ry and $N'_b=400$, $E'_\mathbf{G}=5$~Ry.

\begin{figure}[t]
\centering
\begin{subfigure}[b]{0.3\textwidth}
         \centering
         \includegraphics[width=\textwidth]{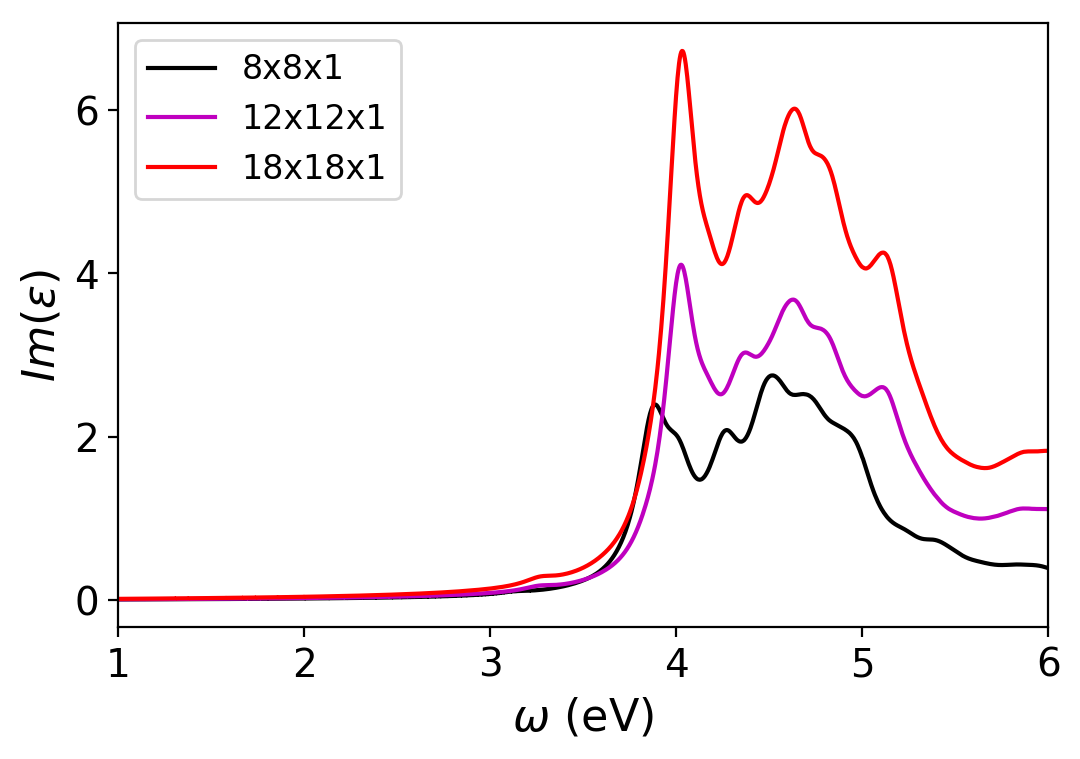}
         \caption{}
         \label{fig:BSE_k}
\end{subfigure}
\hspace{0.2cm}
\begin{subfigure}[b]{0.3\textwidth}
         \centering
         \includegraphics[width=\textwidth]{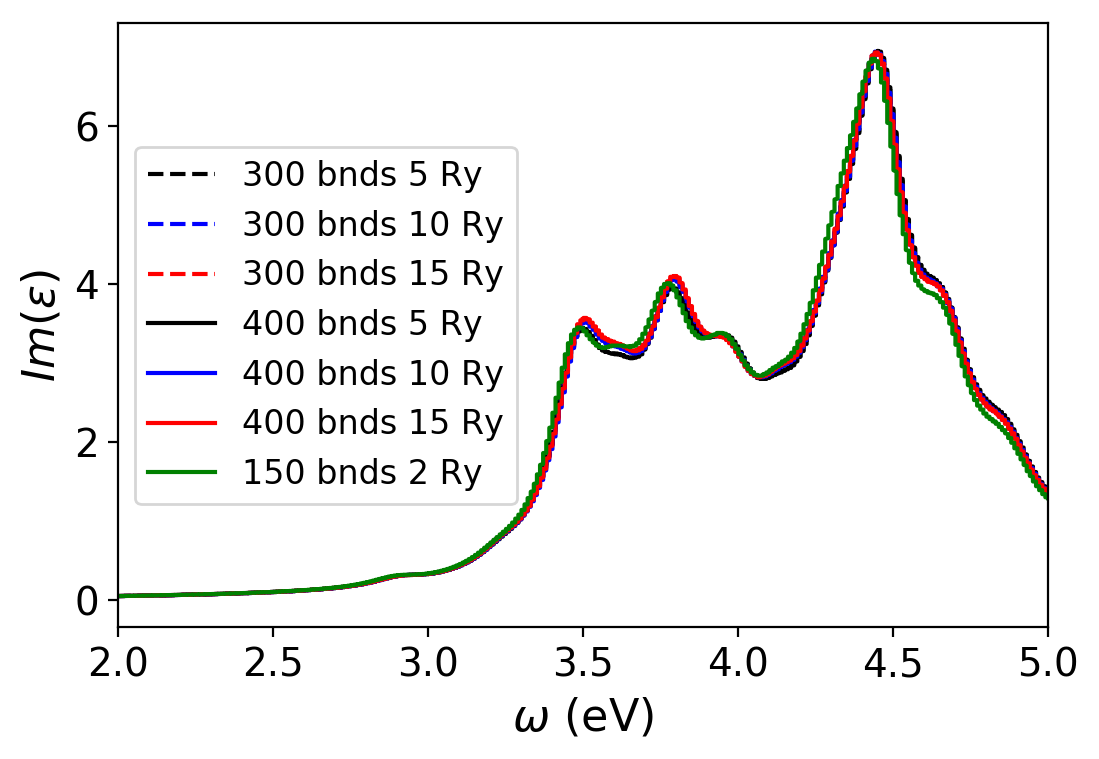}
         \caption{}
         \label{fig:BSE_sex}
\end{subfigure}
\\
\begin{subfigure}[b]{0.3\textwidth}
         \centering
         \includegraphics[width=\textwidth]{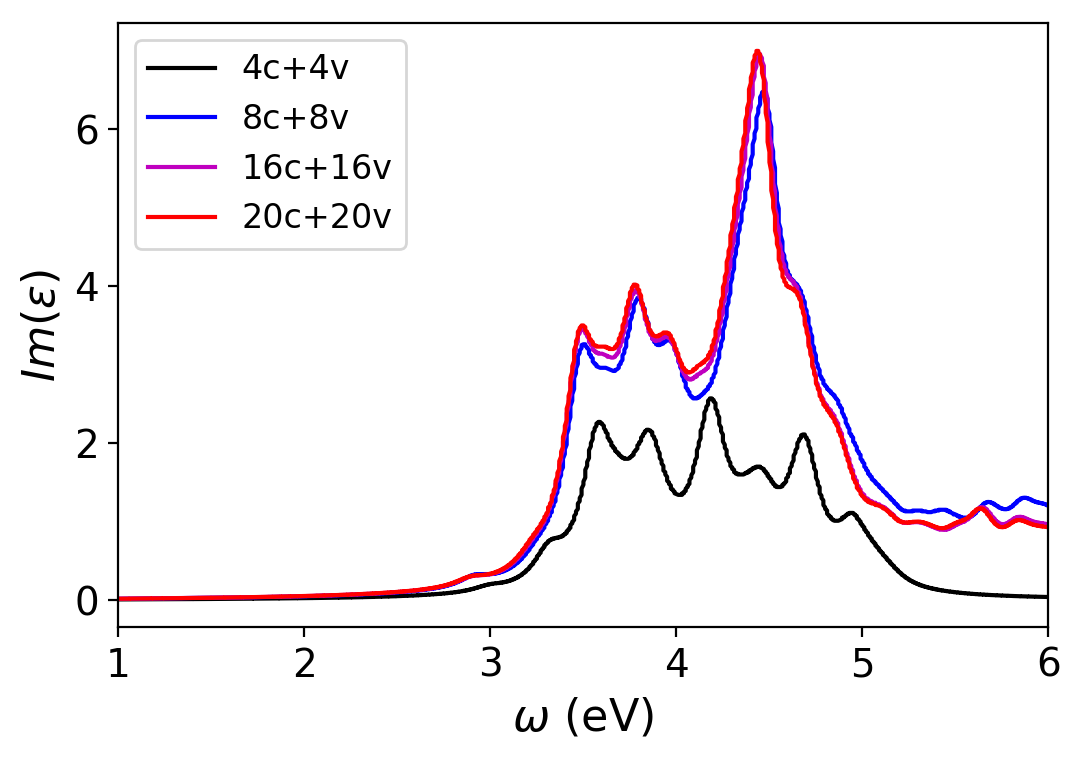}
         \caption{}
         \label{fig:BSE_bnds}
\end{subfigure}
\hspace{0.2cm}
\begin{subfigure}[b]{0.3\textwidth}
         \centering
        \includegraphics[width=\textwidth]{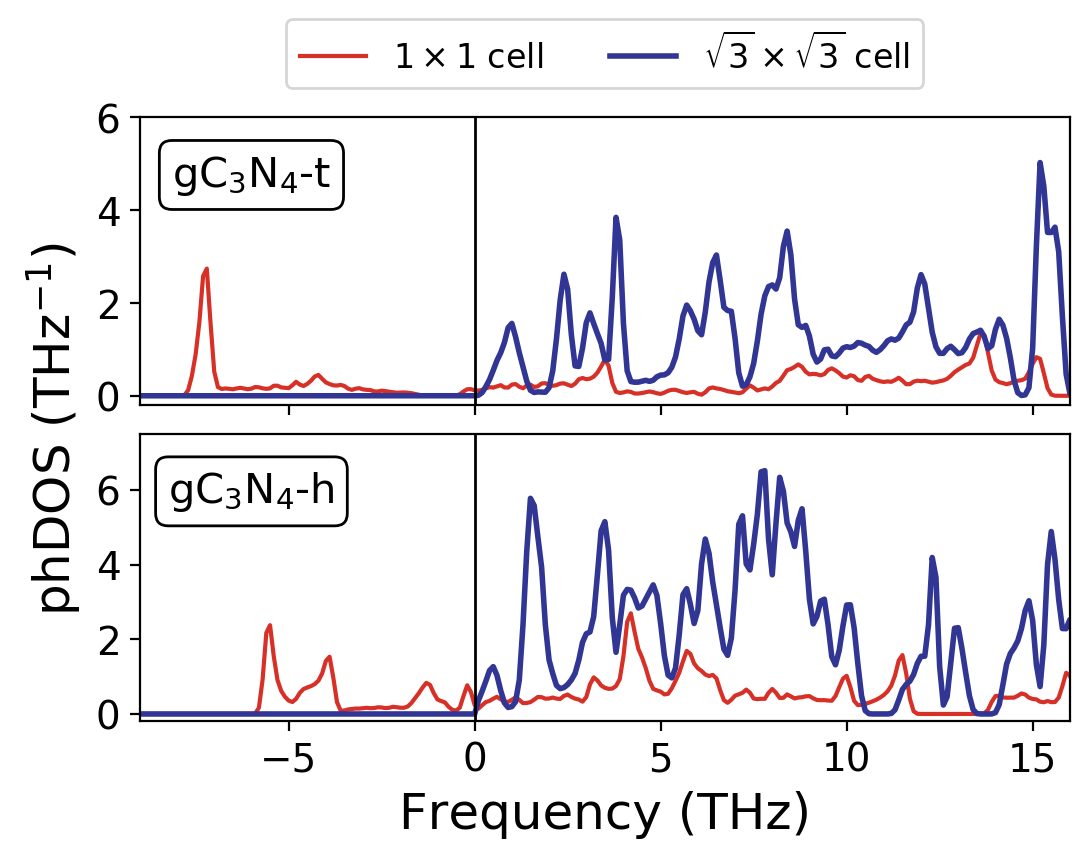}
         \caption{}
	\label{fig:phDOS}
\end{subfigure}
\hfill
\caption{Convergence tests on BSE calculations. (a) Variation of the absorption spectrum on different $k$-point grids. (b) Absorption spectra obtained on the coarse $2\times2\times1$ by varying the number of bands and the energy cutoff of the dielectric screening in the SEX approximation. (c) Variation of the absorption spectrum upon the inclusion in the BSE kernel of different numbers of valence (v) and conduction (c) bands around the electronic bandgap. (d) Phonon Density of States of \gcnt{} (upper panel) and \gcnh{} (lower panel) computed in the $1\times1$ and $\sqrt{3}\times\sqrt{3}$ cells.}
\end{figure}

In BSE calculations, the choice of the $k$-grid is critical. In Fig.~\ref{fig:BSE_k} we report the imaginary part of the macroscopic dielectric function of \gcnt{} computed on different grids. The screening was computed within the static exchange (SEX) approximation, using 150 bands and an energy cutoff of 2~Ry. It is possible to notice that a much denser $18\times18\times1$ grid is necessary to obtain a converged spectrum. By carrying out a series of calculations with increasing number of bands and cutoff on a coarser $2\times2\times1$ grid, Fig.~\ref{fig:BSE_sex}, we verified that the dielectric screening, in the SEX approximation, can be computed by using a smaller number of bands, $N_b^{\rm{BSE}}=150$, and a lower cutoff $E_\mathbf{G}^{\rm{BSE}}=2$~Ry. The convergence of the spectrum also depends on the bands included in the calculation of the BSE kernel. In Fig.~\ref{fig:BSE_bnds} we report the spectra obtained on the coarse $2\times2\times1$ grid, with $N_b^{\rm{BSE}}=150$ and $E_\mathbf{G}^{\rm{BSE}}=2$~Ry, by varying the number of included valence and conduction bands around the bandgap. The inclusion of 8 valence bands and 8 conduction bands around the gap was deemed sufficient for accurately describing the lower part of the spectrum.

In the case of \gcnh{}, a $9\times9\times1$ $k$-grid, $N_b^{\rm{BSE}}=300$, $E_\mathbf{G}^{\rm{BSE}}=2$~Ry and 8 valence + 8 conduction bands were employed.
%
%
%
%
%
\subsubsection*{Density Functional Theory results}
In Tab.~\ref{tab:data} we report the structural and energetic data of the studied \gcnt{} and \gcnh{} cells, obtained with Density Functional Theory. The energy difference per formula unit, $\Delta E / \rm{f.u.}$, is referenced to a unit cell of \gcnt{} in the flat geometry. In Fig.~\ref{fig:phDOS} we report the phonon Density of States of the $1\times1$ and $\sqrt{3}\times\sqrt{3}$ cells of both \gcnt{} (upper panel) and \gcnh{} (lower panel). Soft phonon modes are represented by negative frequencies.
\begin{table}[h]
\begin{tabularx}{\textwidth}{ XX|XXXX }
\multicolumn{2}{c}{structure} & $a$ (\AA) & $b$ (\AA) & $\gamma$ ($^\circ$) & $\Delta E / \mbox{f.u.}$ (eV) \\
\hline
\multirow{5}{*}{\gcnt} & $1\times1$ & 4.786 & 4.786 & 120 & - \\
& $2\times2$ & 9.095 & 9.203 & 118.6 & -0.64 \\
& $3\times3$ & 13.816 & 13.816 & 120 & -0.89 \\
& $4\times4$ & 18.338 & 18.338 & 121 & -0.62 \\
& \sqrtcell{} & 7.975 & 7.975 & 120 & -0.89\\
\hline
\multirow{5}{*}{\gcnh} & $1\times1$ & 7.136 & 7.136 & 120 & -0.36 \\
& $2\times2$ & 13.889 & 13.924 & 120.2 & -0.69 \\
& $3\times3$ & 20.366 & 20.366  & 120 & -1.14 \\
& $4\times4$ & 27.146 & 27.139 & 119.6 & -0.83 \\
& \sqrtcell{} & 11.768 & 11.768 & 120 & -1.14
\end{tabularx}
\caption{Structural data and energetic comparison for \gcnt{} and \gcnh{} in different simulation cells. Here $\gamma$ denotes the angle between the in-plane lattice vectors $a$ and $b$.}
\label{tab:data}
\end{table}
\clearpage
\subsubsection*{Wavefunctions of selected states of \gcnh{}}
\begin{figure}[h]
\centering
\begin{subfigure}[b]{0.3\textwidth}
         \centering
         \includegraphics[width=\textwidth]{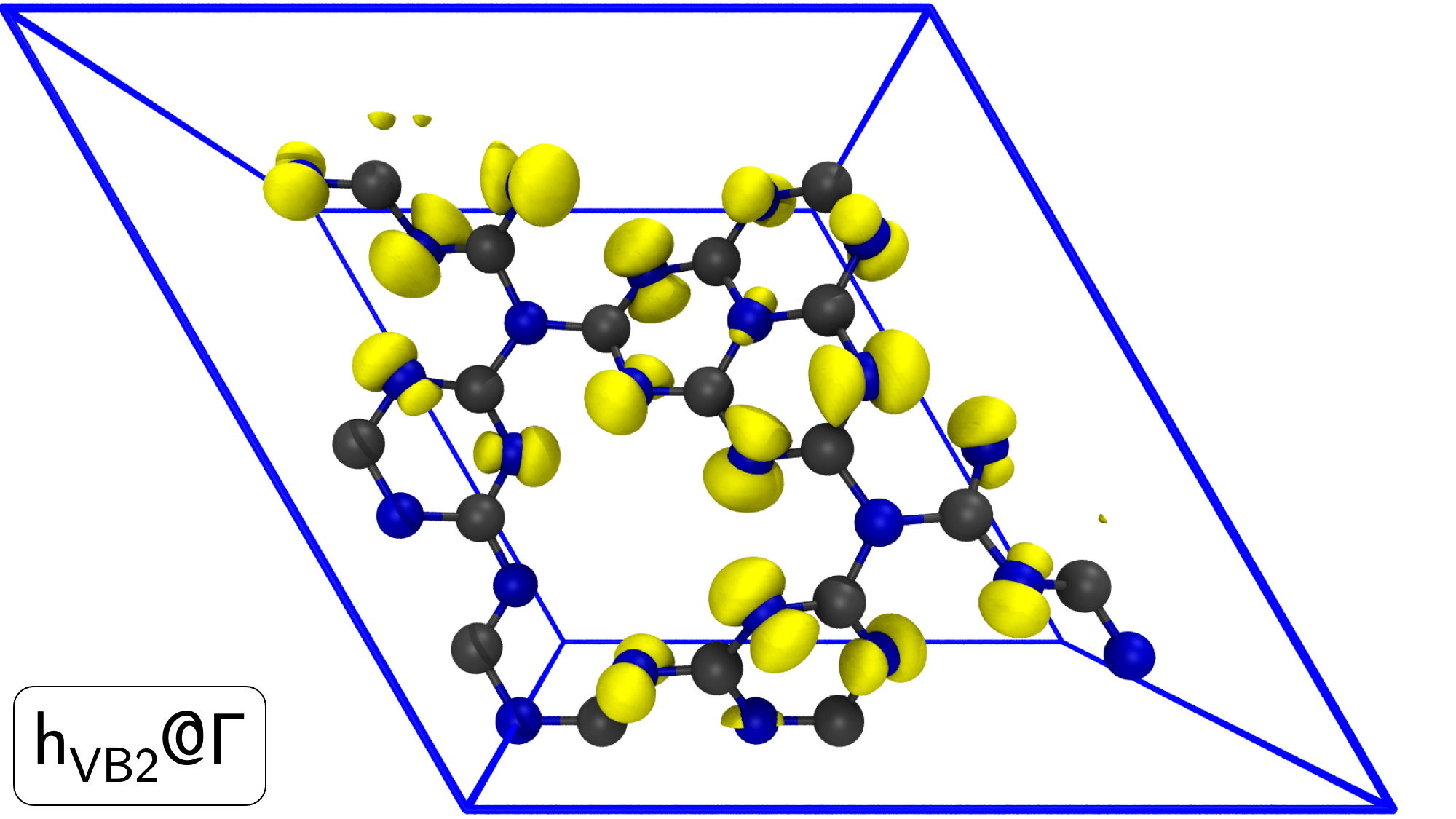}
\end{subfigure}
\hspace{.3cm}
\begin{subfigure}[b]{0.3\textwidth}
         \centering
         \includegraphics[width=\textwidth]{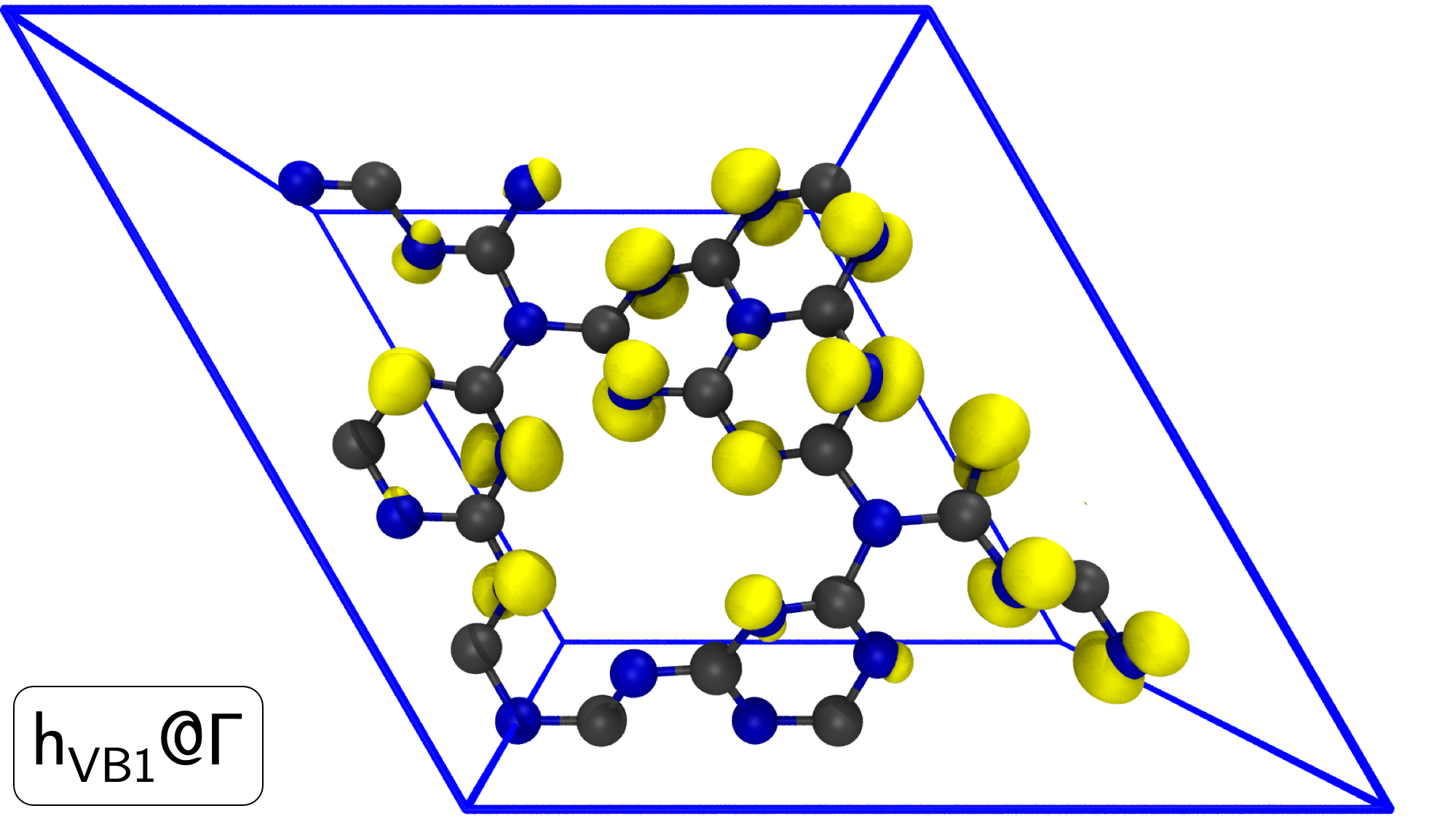}
\end{subfigure}\\
\vspace{.8cm}
\begin{subfigure}[b]{0.3\textwidth}
         \centering
         \includegraphics[width=\textwidth]{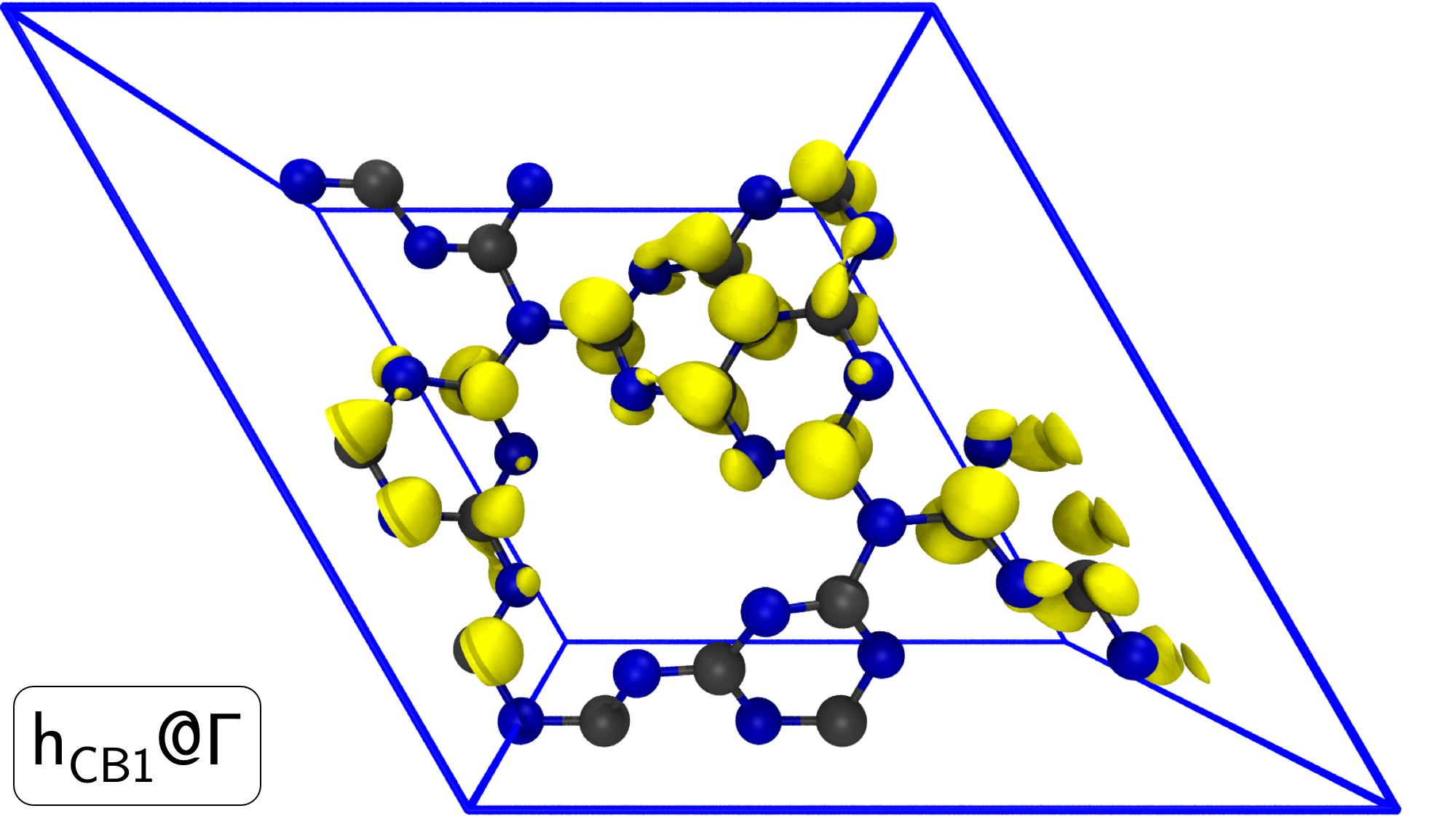}
\end{subfigure}
\hspace{.3cm}
\begin{subfigure}[b]{0.3\textwidth}
         \centering
         \includegraphics[width=\textwidth]{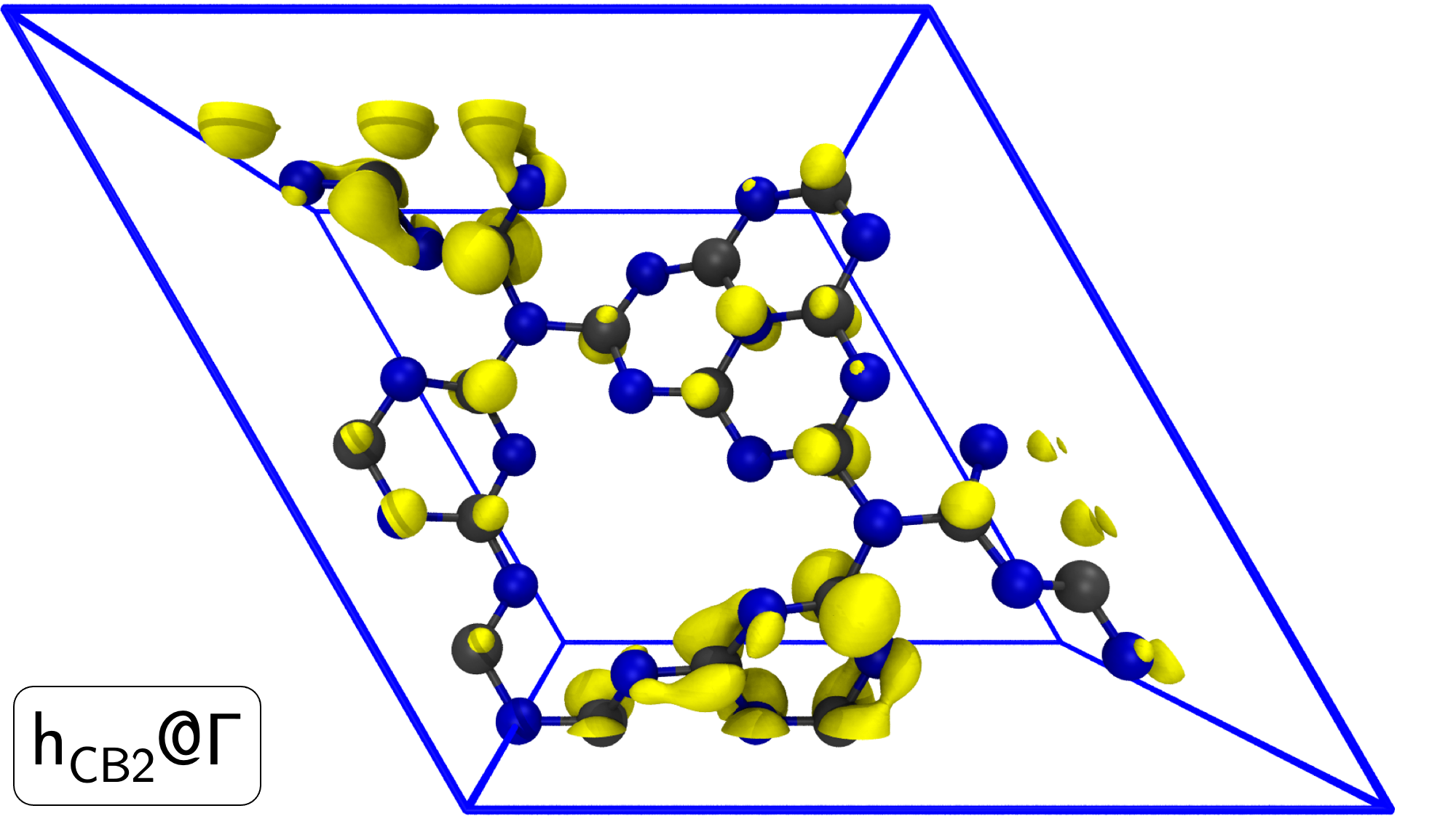}
\end{subfigure}\\
\vspace{.8cm}
\begin{subfigure}[b]{0.3\textwidth}
         \centering
         \includegraphics[width=\textwidth]{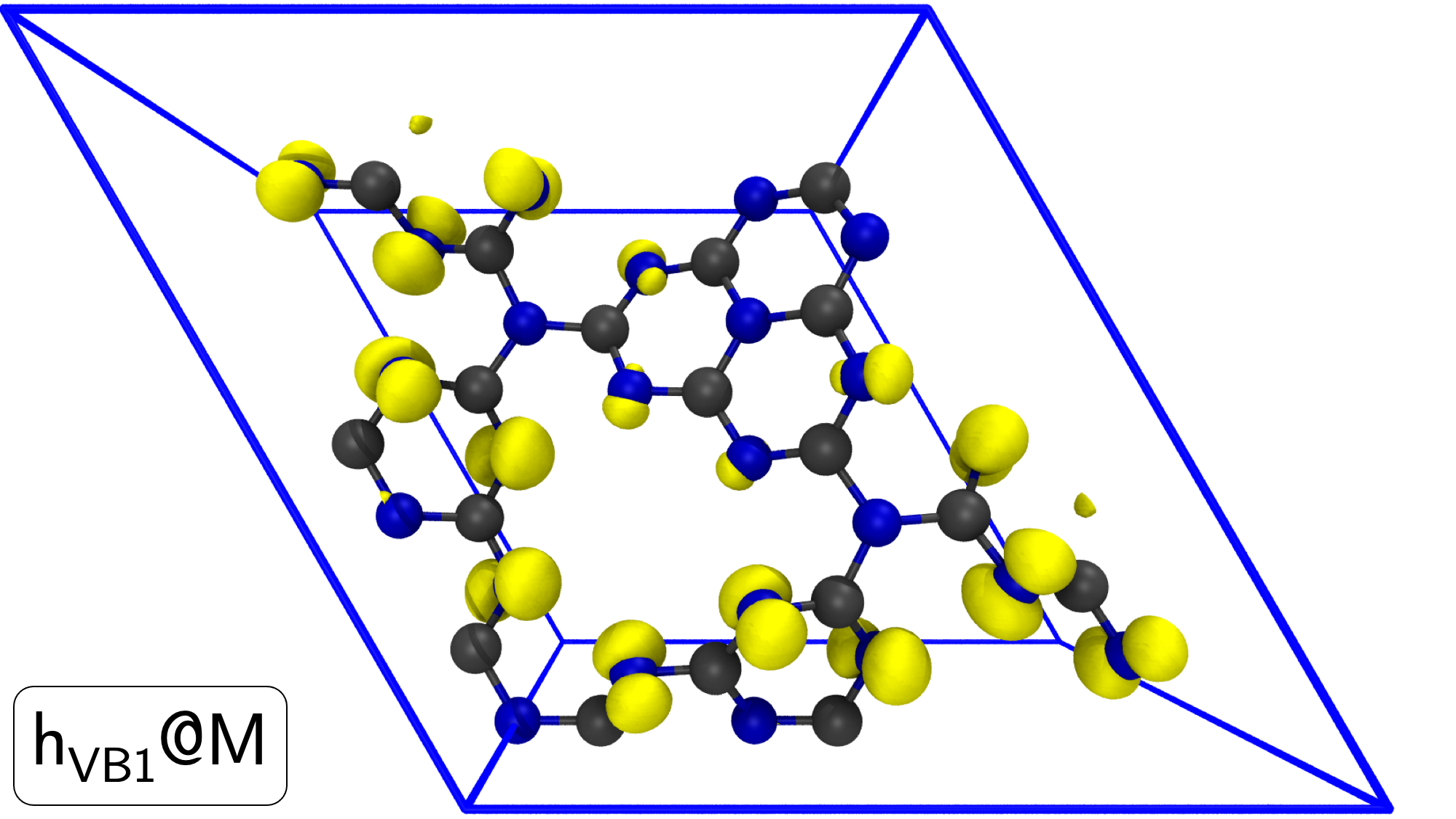}
\end{subfigure}
\hspace{.3cm}
\begin{subfigure}[b]{0.3\textwidth}
         \centering
         \includegraphics[width=\textwidth]{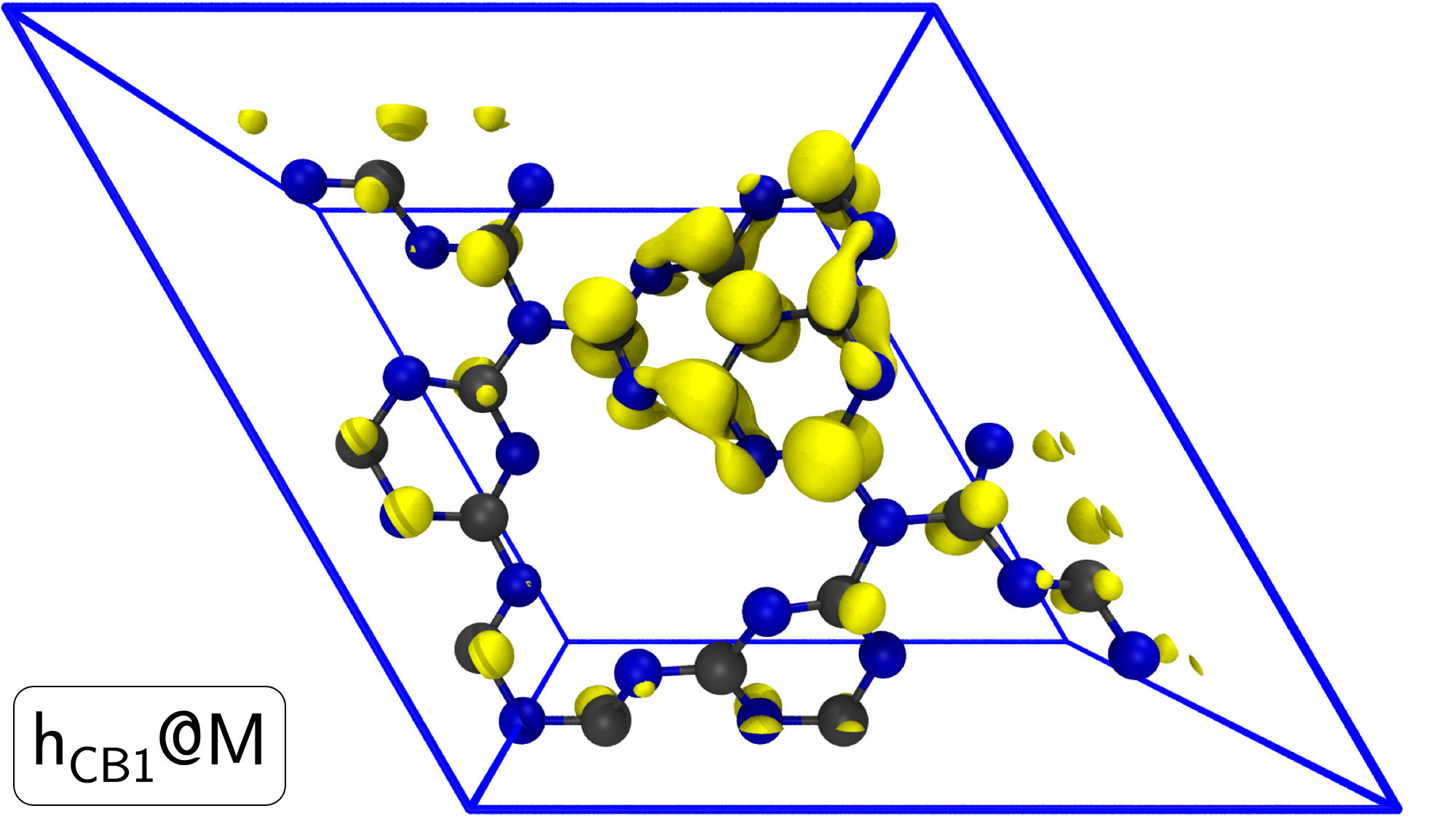}
\end{subfigure}\\
\vspace{.5cm}
\caption{Isosurfaces of the wavefunctions of the states involved in the most significant transitions contributing to \gcnh{} excitons.}
\label{fig:wf}
\end{figure}

In Fig.~\ref{fig:wf} we report the square modulus of the wavefunctions of the \gcnh{} electronic states involved in the optical transitions mentioned in the text.\\
\bigskip

\end{document}